\documentclass[12pt]{article}
\usepackage{graphicx}% Include figure files
\usepackage{amssymb}
\usepackage{amsmath}
\newcommand{\vev}[1]{{\langle #1 \rangle}}

\newcommand{\MeV}{\mbox{~MeV}}
\newcommand{\GeV}{\mbox{~GeV}}

\newcommand{\ie}{{\it i.e.}}

\newcommand{\eqn}[1]{&\hspace{-0.6em}#1\hspace{-0.6em}&}
\begin{document}
\baselineskip 0.6cm
%
%%%%%%%%%%%%%%%%%%%%%%%%%%%%%%%%%%%%%%%%%%%%%%%%%%%%%%%%%%%%%%%%%%%%
%%%%%  Title Page  %%%%%%%%%%%%%%%%%%%%%%%%%%%%%%%%%%%%%%%%%%%%%%%%%
\begin{titlepage}
\begin{center}

%%%%%%%% Preprint #
\begin{flushright}
%DESY 01-142\\
\end{flushright}

\vskip 2cm

%%%%%%%% Title
{\Large \bf 
Kinetic Equations for Baryogenesis \\[1ex]
via
Sterile Neutrino Oscillation
}

\vskip 1.2cm

%%%%%%%% Authors
{\large 
Takehiko Asaka$^{1,2}$, Shintaro Eijima$^{2,3}$ and Hiroyuki Ishida$^{2,3}$
}

\vskip 0.4cm

%%%%%%%% Addresses
$^1${\em
  Department of Physics, Niigata University, Niigata 950-2181, Japan
}

$^2${\em
  Max-Planck-Institut f\"ur Kernphysik,
  Postfach 103980, 69029 Heidelberg, Germany
}

$^3${\em
  Graduate School of Science and Technology, Niigata University, Niigata 950-2181, Japan
}

\vskip 0.2cm

%%%%%%%% Date
%(\today)
(December 23, 2011)

\vskip 2cm

\vskip .5in
%%%%%%%%%%%%%%%% Abstract %%%%%%%%%%%%%%%%%%%%%%%%%%%%%%%%%%%%%%%%%%
\begin{abstract}
  We investigate baryogenesis in the $\nu$MSM (neutrino Minimal
  Standard Model), which is the MSM extended by three right-handed
  neutrinos with masses below the electroweak scale.  The baryon
  asymmetry of the universe can be generated by the mechanism via
  flavor oscillation of right-handed (sterile) neutrinos which are
  responsible to masses of active neutrinos confirmed by various
  experiments.  We present the kinetic equations for the matrix of
  densities of leptons which describe the generation of asymmetries.
  Especially, the momentum dependence of the matrix of densities is
  taken into account.  By solving these equations numerically, it is
  found that the momentum distribution is significantly distorted from
  the equilibrium one, since the production for the modes with lower
  momenta $k \ll T$ ($T$ is the temperature of the universe) is
  enhanced, while suppressed for higher modes.  As a result, the most
  important mode for the yields of sterile neutrinos as well as the
  baryon asymmetry is $k \simeq 2 T$, which is smaller than $\langle k
  \rangle$ inferred from the thermal average.  The comparison with the
  previous works is also discussed.
\end{abstract}
%%%%%%%%%%%%%%%% Abstract %%%%%%%%%%%%%%%%%%%%%%%%%%%%%%%%%%%%%%%%%%
%%%%%%%%%%%%%%%%%%%%%%%%%%%%%%%%%%%%%%%%%%%%%%%%%%%%%%%%%%%%%%%%%%%%
\end{center}
\end{titlepage}
%%%%%  Title Page  %%%%%%%%%%%%%%%%%%%%%%%%%%%%%%%%%%%%%%%%%%%%%%%%%
%%%%%%%%%%%%%%%%%%%%%%%%%%%%%%%%%%%%%%%%%%%%%%%%%%%%%%%%%%%%%%%%%%%%
\renewcommand{\thefootnote}{\#\arabic{footnote}} 
\setcounter{footnote}{0}
%\clearpage
%%%%%%%%%%%%%%%%%%%%%%%%%%%%%%%%%%%%%%%%%%%%%%%%%%%%%%%%%%%%%%%%%%%%
%%%%% ** Text ** %%%%%%%%%%%%%%%%%%%%%%%%%%%%%%%%%%%%%%%%%%%%%%%%%%%
%%%%%%%%%%%%%%%%%%%%%%%%%%%%%%%%%%%%%%%%%%%%%%%%%%%%%%%%%%%%%%%%%%%%
%
%
%%%%%%%%%%%%%%%%%%%%%%%%%%%%%%%%%%%%%%%%%%%%%%%%%%%%%%%%%%%%%%%%%%%%
%%%%%%%%%%%%%%%%%%%%%%%%%%%%%%%%%%%%%%%%%%%%%%%%%%%%%%%%%%%%%%%%%%%%
\section{Introduction}
\label{sec:Introduction}
%%%%%%%%%%%%%%%%%%%%%%%%%%%%%%%%%%%%%%%%%%%%%%%%%%%%%%%%%%%%%%%%%%%%
%%%%%%%%%%%%%%%%%%%%%%%%%%%%%%%%%%%%%%%%%%%%%%%%%%%%%%%%%%%%%%%%%%%%
The baryon asymmetry of the universe (BAU) is one of the most puzzling
problems in particle physics and cosmology.  It is certain that the
Minimal Standard Model (MSM) cannot account for its origin.
There have so far been proposed various scenarios of baryogenesis by
considering physics beyond the MSM. (See, for example, a recent review
\cite{Riotto:1999yt}.)  Among them leptogenesis~\cite{Fukugita:1986hr}
by superheavy right-handed neutrinos is one of the most motivated
scenarios.  This is because these new particles allow us to give
non-zero masses to neutrinos which have been confirmed in many
oscillation experiments.  Being singlet under the MSM gauge group they
can obtain Majorana masses which are completely independent on the
electroweak scale.  Right-handed neutrinos having superheavy masses
then can generate the lepton asymmetry by their decays which can be a
source of the BAU.  In addition, the observed smallness of active
neutrino masses is naturally explained by such fermions
through the seesaw mechanism~\cite{Seesaw}.  The required masses
are so heavy that it is impossible to directly test these
new particles in near future experiments.

Akhmedov, Rubakov and Smirnov have proposed~\cite{Akhmedov:1998qx}
another attractive scenario of baryogenesis by using right-handed
neutrinos (mechanism via neutrino oscillation).  One of the most
important features is that the mechanism works when their masses are
smaller than the electroweak scale, which are within the reach of
current experiments.  The flavor oscillation among right-handed
neutrinos in the early universe leads to the separation of the lepton
number into right-handed neutrinos and left-handed leptons.  The
sphaleron process then converts the asymmetry of the left-handed
leptons into the baryon asymmetry for high
temperatures~\cite{Kuzmin:1985mm}.

This mechanism can be realized in a simple and attractive framework
called as the neutrino MSM ($\nu$MSM)~\cite{Asaka:2005an,Asaka:2005pn}
in which three right-handed neutrinos are introduced with Majorana
masses below the electroweak scale.  Because of the very suppressed
neutrino Yukawa interaction the seesaw mechanism is still effective,
which leads to three active neutrinos $\nu_i$ ($i=1,2,3$) and three
sterile neutrinos $N_I$ ($I=1,2,3$) as mass eigenstates.  The former
ones are responsible to the neutrino oscillations observed in
experiments.  The latter ones give the solutions to the cosmological
problems.  The lightest sterile neutrino $N_1$ with mass $\sim 10$ keV
can be a candidate of dark matter, while the rest two $N_2$ and $N_3$
with masses $\sim 1$ GeV can generate the BAU through the mechanism
above. (For a review see Ref.~\cite{Boyarsky:2009ix}.)  Furthermore,
the non-minimal coupling of Higgs field to gravity allows to realize
the cosmic inflation~\cite{Bezrukov:2007ep}.

Various aspects of baryogenesis in the $\nu$MSM has been studied until
now~\cite{Asaka:2005pn,Shaposhnikov:2008pf,Asaka:2010kk,Canetti:2010aw}.
The generation of the asymmetry is described by the matrix of
densities~\cite{Dolgov:1980cq,Barbieri:1990vx,Sigl:1992fn} of sterile
(right-handed) neutrinos and active (left-handed) leptons.  The
previous works were based on the kinetic equations for the matrix of
densities in Ref.~\cite{Asaka:2005pn}, where the terms describing the
transfer of asymmetries between sterile neutrinos and active leptons
are added to the original ones in \cite{Akhmedov:1998qx}.  As shown in
\cite{Asaka:2005pn}, such terms are essential to generate enough
amount of the BAU in the $\nu$MSM.  In spite of their significance
such terms are introduced by a heuristic approach.

The main motivation of this article is to improve the estimation of
the BAU in the $\nu$MSM.  One of the most unsatisfactory points in the
previous works is that the evolution of the matrix of densities has
been traced only by an approximate way.  Namely, the kinetic equations
analyzed previously are for the typical, single mode with momentum $k
\sim T$ ($T$ is the temperature of the universe), which is expected to
give a dominant contribution to the asymmetry.  By extrapolating the
result of such a mode to other modes the asymmetry in the number
density is estimated.

In this article, thus, we derive the kinetic equations in which the
momentum dependence of the matrix of densities is fully taken into
account.  For this purpose we shall re-evaluate the destruction and
production rates of sterile neutrinos as well as active leptons paying
attention to the two points.  One is the momentum dependence of the
particle of interest.  The other is the corrections coming from the
fact that the states in the destruction and production processes may
differ from the thermal equilibrium states (\ie, the deviations from
the thermal equilibrium for sterile neutrinos and the chemical
potentials for active leptons).  As we will show later, the kinetic
equations are written as simultaneous integrodifferential equations
for the matrices of densities of sterile neutrinos and active leptons.

Interestingly, our kinetic equations include the terms connecting
between sterile and active sectors, which is crucial for the
baryogenesis in the $\nu$MSM as mentioned before.  It will be shown
that such terms arise automatically as the corrections to the
destruction and production rates owing to the deviations from the
equilibrium states in the scattering processes.  It turns out that the
coefficients of these terms are different from~\cite{Asaka:2005pn}
and, in addition, there appears a new type of terms which couples
sterile neutrinos to active anti-leptons (rather than active leptons).

This paper is organized as follows: In Sec.~\ref{sec:BGinNuMSM} we
briefly review the framework of the analysis and the baryogenesis
scenario in the $\nu$MSM.  In the Sec.~\ref{sec:KE}, we derive the
kinetic equations which describe the generation of the baryon
asymmetry paying attention to the momentum dependence in the matrices
of densities.  Sec.~\ref{sec:NS} is devoted to study the numerical
solutions to the obtained equations.  We show the mode by mode
evolution of the matrices of densities and their momentum
distributions.  The comparison with the previous works is done in
Sec.~\ref{sec:Comparison}.  We perform the quantitative comparison of
the yields of the BAU, but also clarify the differences between the
kinetic equations in the literature.  We conclude in
Sec.~\ref{sec:Conc}.

%%%%%%%%%%%%%%%%%%%%%%%%%%%%%%%%%%%%%%%%%%%%%%%%%%%%%%%%%%%%%%%%%%%%
%%%%%%%%%%%%%%%%%%%%%%%%%%%%%%%%%%%%%%%%%%%%%%%%%%%%%%%%%%%%%%%%%%%%
\section{Baryogenesis in the $\nu$MSM}
\label{sec:BGinNuMSM}
%%%%%%%%%%%%%%%%%%%%%%%%%%%%%%%%%%%%%%%%%%%%%%%%%%%%%%%%%%%%%%%%%%%%
%%%%%%%%%%%%%%%%%%%%%%%%%%%%%%%%%%%%%%%%%%%%%%%%%%%%%%%%%%%%%%%%%%%%
At the beginning we review the $\nu$MSM, which is the MSM extended by
three right-handed neutrinos $\nu_R{}_ I$ ($I = 1,2,3$) with
Lagrangian
\begin{equation}
  \label{eq:L_nuMSM}
  {\cal L}_{\nu{\rm MSM}} =
  {\cal L}_{\rm MSM} +
  i \, \overline{\nu_R{}_I} \, \gamma^\mu \, \partial_\mu \, \nu_R{}_I
  -
  \Bigl(
  F_{\alpha I} \, \overline{L}_\alpha \, \Phi \, \nu_R{}_I
  + \frac{M_I}{2} \, \overline{\nu_R{}_I^c} \, \nu_R{}_I 
  + h.c.
  \Bigr)
\,.
\end{equation}
where ${\cal L}_{\rm MSM}$ is the MSM Lagrangian, $F_\alpha{}_I$ are
neutrino Yukawa coupling constants, and $\Phi$ and $L_\alpha$ ($\alpha
= e, \nu, \tau$) are Higgs and lepton weak-doublets, respectively.
The Majorana masses of right-handed neutrinos are denoted by $M_I$.
After the electroweak symmetry breaking neutrinos also obtain the
Dirac masses, $[M_D]_\alpha{}_I = F_\alpha{}_I \vev{\Phi}$
($\vev{\Phi}$ is a vacuum expectation value of the Higgs field), and
tiny neutrino masses can be explained by the seesaw mechanism if
$\left| [M_D]_\alpha{}_I \right| \ll M_I$ is satisfied.  In this case,
mass eigenstates of neutrinos are consist of active neutrinos $\nu_i$
($i = 1,2,3$) and sterile neutrinos $N_I$ ($I = 1,2,3$).  Active
neutrinos are the mass eigenstates of ordinary neutrinos and sterile
neutrinos are almost the right-handed states $N_I \simeq \nu_R{}_I$
with masses given by the Majorana masses $M_I$ approximately.

On the other hand, when we discuss baryogenesis, the temperatures of
interest is higher than the electroweak scale, and we can neglect all
the masses of the MSM particles and also the mixing between active and
sterile neutrinos.  In this case we can treat active and sterile
leptons as left-handed leptons and right-handed neutrinos,
respectively.

Among three sterile neutrinos, the lightest one $N_1$ is a candidate
of dark matter.  In this dark matter scenario, the coupling constants
of neutrino Yukawa interaction of $N_1$ are required to be very
small. (See, for example, a review~\cite{Boyarsky:2009ix}.)  As a
result, $N_1$ gives the negligible contributions to the seesaw mass
matrix of active neutrinos as well as the BAU, and then two sterile
neutrinos $N_2$ and $N_3$ are responsible to these two phenomena.

In the $\nu$MSM the sufficient amount of baryon asymmetry can be
generated~\cite{Asaka:2005pn} by the mechanism via neutrino
oscillation~\cite{Akhmedov:1998qx}.  The flavor oscillation between
sterile (right-handed) neutrinos $N_2$ and $N_3$ are induced by the
medium effect, which can generate the asymmetries in leptons together
with the CP violation in neutrino Yukawa matrix.  The asymmetry in
active (left-handed) leptons is then transformed into the baryon
asymmetry due to the $B+L$ breaking sphaleron process for high
temperatures $T> T_W = {\cal O}(10^2)$ GeV~\cite{Kuzmin:1985mm}.
Notice that sterile neutrinos is possible to be out of thermal
equilibrium due to the smallness of neutrino Yukawa coupling
constants.  If this is the case, all the conditions for successful
baryogenesis~\cite{Sakharov:1967dj} can be satisfied.

In order to describe a series of the processes generating asymmetries,
we have to deal with the coherent evolution of sterile neutrinos with
the flavor oscillation, and also the incoherent scattering processes
with surrounding medium for the destruction and production of sterile
neutrinos.  We then use the formulation with the matrix of
densities~\cite{Dolgov:1980cq,Barbieri:1990vx,Sigl:1992fn}.  The
matrix of interest here is one in the flavor space consisting of
sterile neutrinos $N_2$ and $N_3$ with positive helicities, $\bar N_2$
and $\bar N_3$ with negative helicities, active leptons $L_e$, $L_\mu$
and $L_\tau$, and their anti-particles $\bar L_e$, $\bar L_\mu$ and
$\bar L_\tau$, which leads to a $10 \times 10$ matrix.  The diagonal
elements of this matrix are nothing but the usual occupation numbers,
and the off-diagonal elements contain correlations of the flavor
mixings.

In the situation under consideration, however, it can be simplified as
described in Ref.~\cite{Asaka:2005pn}.  First of all, the conservation
of the total lepton number holds for the temperatures of baryogenesis
($T> T_W$), since Majorana masses $M_I$ for sterile neutrinos are
sufficiently small in the $\nu$MSM.  We can thus neglect all the
elements which break the lepton number.  Second, active leptons
possess gauge interactions and Yukawa interactions with right-handed
charged leptons in addition to neutrino Yukawa interactions, which
induce large energy gaps between active and sterile states through the
thermal effect. Then, the transitions between active and sterile
leptons are highly suppressed, and the elements corresponding to the
mixing between $L_\alpha$ and $N_I$ (and also $\bar L_\alpha$ and
$\bar N_I$) are neglected.  At this stage, the matrix of densities in
the system is decomposed into two $2 \times 2$ matrices $\rho_N$ and
$\rho_{\bar N}$ for $N_{2,3}$ and $\bar N_{2,3}$, and two $3 \times 3$
matrices $\rho_L$ and $\rho_{\bar L}$ for active leptons
$L_{e,\mu,\tau}$ and $\bar L_{e,\mu,\tau}$.  As in
Ref.~\cite{Asaka:2010kk}, $\rho_L$ is considered to be $\rho_L =
\rho_\nu + \rho_e = N_D \rho_\nu$, \ie, the sum of two contributions
of lepton doublet.  Finally, the off-diagonal elements of $\rho_L$ and
$\rho_{\bar L}$ can be neglected for the temperatures of interest.
This is because the flavor transitions among active leptons are also
suppressed due to the medium effects induced by the Yukawa
interactions of charged leptons with hierarchical coupling constants.
Note that active leptons can maintain the kinetic equilibrium due to
the rapid interactions with medium, and hence their diagonal matrices
of densities can be expressed by using the dimensionless chemical
potentials $\mu_{\nu_\alpha}$ (the ordinary chemical potential divided
by temperature) as
\begin{eqnarray}
  \label{eq:RHOL}
  \rho_{L} (k) = N_D \rho^{\rm eq}(k) \, A \,,~~~
  \rho_{\bar L} (k) = N_D \rho^{\rm eq}(k) \, A^{-1} \,,
\end{eqnarray}
where $N_D=2$ is an SU(2) factor, $A = \mbox{diag}(e^{\mu_{\nu_e}},
e^{\mu_{\nu_\mu}}, e^{\mu_{\nu_\tau}})$ and $\rho^{\rm eq}(k)$ is the
equilibrium distribution function.  Throughout this analysis we
apply the approximation of the Boltzmann statistics in which
$\rho^{\rm eq}(k) = e^{-k/T}$.

The evolution of the system is now described by two $2 \times 2$
matrices of densities $\rho_N(k)$ and $\rho_{\bar N}(k)$, and three
differences of the distribution functions $[\rho_L (k) - \rho_{\bar L}
(k)]_{\alpha \alpha}$ (or three chemical potentials
$\mu_{\nu_{\alpha}}$).  The previous works on baryogenesis via
neutrino oscillations have been based on the kinetic equations for
these variables given in Ref.~\cite{Akhmedov:1998qx} or
\cite{Asaka:2005pn}.  The original work~\cite{Akhmedov:1998qx} has
considered the kinetic equations for $\rho_N$ and $\rho_{\bar N}$
without including the effects from active leptons.  As pointed out in
Ref.~\cite{Asaka:2005pn}, their equations give too small baryon
asymmetry to account for the observed value in the $\nu$MSM. This is
because the lightest sterile neutrino $N_1$ is a dark matter candidate
and essentially only two sterile neutrinos $N_2$ and $N_3$ participate
in baryogenesis.  Furthermore, Ref.~\cite{Asaka:2005pn} has introduced
the terms which connect active and sterile sectors, and solved the
kinetic equations for both $\rho_{N, \bar N}$ and $\rho_{L, \bar L}$.
It has been shown that the additional terms boost the generation of
the asymmetry and hence are vital for the successful baryogenesis in
the $\nu$MSM.

There are, however, various aspects to be improved in the previous
works towards the precise prediction of the baryon asymmetry in the
$\nu$MSM.  One of the most important points is to include correctly
the momentum dependence in the matrices of densities $\rho_N(k)$ and
$\rho_{\bar N} (k)$.  So far, it has been analyzed only the evolution
of the single mode with a typical momentum $k \sim T$, and estimated
the asymmetries in number densities by the approximation that the
occupation numbers are proportional to the equilibrium one $\rho^{\rm
  eq}(k)$.  To get rid of this uncertainty, therefore, we will write
down the kinetic equations of the matrices of densities for all the
modes and solve them numerically taking care of the expansion of the
universe.  It will be shown later that the momentum distributions of
$\rho_{N, \bar N}$ are distorted significantly from $\rho^{\rm eq}$.

%%%%%%%%%%%%%%%%%%%%%%%%%%%%%%%%%%%%%%%%%%%%%%%%%%%%%%%%%%%%%%%%%%%%
%%%%%%%%%%%%%%%%%%%%%%%%%%%%%%%%%%%%%%%%%%%%%%%%%%%%%%%%%%%%%%%%%%%%
\section{Kinetic Equations}
\label{sec:KE}
%%%%%%%%%%%%%%%%%%%%%%%%%%%%%%%%%%%%%%%%%%%%%%%%%%%%%%%%%%%%%%%%%%%%
%%%%%%%%%%%%%%%%%%%%%%%%%%%%%%%%%%%%%%%%%%%%%%%%%%%%%%%%%%%%%%%%%%%%
Now we are at the position to derive the kinetic equations for
$\rho_{N, \bar N}$ and $\rho_{L, \bar L}$.  
First of all, let us  consider the time evolution of $\rho_N$.
Our construction is based on Ref.~\cite{Sigl:1992fn}
(as in \cite{Akhmedov:1998qx,Asaka:2005pn}) and starts with
\begin{eqnarray}
  \label{eq:RHON}
  \frac{d \rho_N(k_N)}{dt}
  =
  - i \big[H_N(k_N) ,\, \rho_N(k_N)  \big]
  - \frac{1}{2} \big\{ \Gamma_N^{d}(k_N) , \, \rho_N(k_N) \big\}
  + \frac{1}{2} \big\{ \Gamma_N^{p}(k_N) , \, 
  \mathbf{1} - \rho_N(k_N) \big\}\,,
\end{eqnarray}
where $\mathbf{1}$ denotes the unit matrix with appropriate
dimensions.%
\footnote{
We have neglected the
non-linear effects of $\rho_N$ since the interaction rates between
sterile neutrinos are sufficiently small. 
Otherwise, see Ref.~\cite{Dasgupta:2009mg,Duan:2010bg}.
}  
$H_N$ is the effective Hamiltonian, $H_N = H_N^0 + V_N$,
where the free part is $[H_N^0(k_N)]_{IJ} = E_{N_I} \delta_{IJ}$ with
$E_{N_I} = \sqrt{k_N^2+M_I^2}$ and $V_N$ is the effective potential
induced by the medium effects.  $\Gamma_N^d$ and $\Gamma_N^p$ are the
destruction and production rates of $N_I$.  From now on we shall apply
the approximation of the Boltzmann statistics and replace the third
term of Eq.~(\ref{eq:RHON}) as $\frac{1}{2} \{ \Gamma_N^{p} , \,
\mathbf{1} - \rho_N \} \to \Gamma_N^p$.  

%%%%%%%%%%%%%%%%%%%%%%%%%%%%%%%%%%%%%%%%%%%%%%%%%%%%%%%%%%%%%%%%%%%%
%%%%% ** Figure ** %%%%%%%%%%%%%%%%%%%%%%%%%%%%%%%%%%%%%%%%%%%%%%%%%
\begin{figure}[t]
  \centerline{
  \includegraphics[scale=0.6]{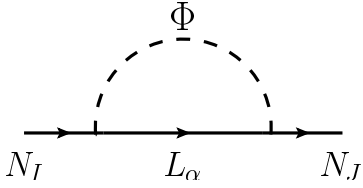}
  }%
  \caption{\it
    Feynman diagram for self energy of sterile neutrinos.
  }
  \label{fig:Self}
\end{figure}
%%%%%%%%%%%%%%%%%%%%%%%%%%%%%%%%%%%%%%%%%%%%%%%%%%%%%%%%%%%%%%%%%%%%
The first term of Eq.~(\ref{eq:RHON}) describes the coherent evolution
of $\rho_N$ and the oscillation of sterile neutrinos occurs due to the
off-diagonal elements of $V_N$, which is essential for baryogenesis
under consideration.  It is found from the self energy for sterile
neutrinos at finite temperatures in Fig.~\ref{fig:Self} that the
effective potential for the mode $k=k_N$ is given
by~\cite{Weldon:1982bn}
\begin{eqnarray}
  \big[ V_N (k_N) \big]_{IJ}
  = \frac{N_D T^2}{16 \, k_N} \,\big[ F^\dagger F \big]_{IJ} 
  \,,
%  ~~~~
%  \big[ V_{\bar N} (k) \big]_{IJ}
%  \eqn{=} \frac{N_D T^2}{16 \, k} \,\big[ F^T F^\ast \big]_{IJ} 
%  \,,
\end{eqnarray}
where we disregard the correction to $V_N$ from the asymmetries in
active leptons.%
\footnote{ We have numerically confirmed that the change of the final
  baryon asymmetry by this effect is negligibly small.  } 

In the estimation of $V_N$ (as well as $\Gamma^{d,p}_N$ below) 
all masses including $M_I$ are neglected
since they are irrelevant for temperatures of interest.  
(Note, however, that we keep $M_I$ in
$H_N^0$ because they are crucial for the oscillation of sterile
neutrinos.)  Further, we first calculate them in the basis where 
neutrino Yukawa matrix is diagonal, and then find the expression in
the original basis shown in Eq.~(\ref{eq:L_nuMSM}).

%%%%%%%%%%%%%%%%%%%%%%%%%%%%%%%%%%%%%%%%%%%%%%%%%%%%%%%%%%%%%%%%%%%%
%%%%% ** Figure ** %%%%%%%%%%%%%%%%%%%%%%%%%%%%%%%%%%%%%%%%%%%%%%%%%
\begin{figure}[t]
  \centerline{
  \includegraphics[scale=0.6]{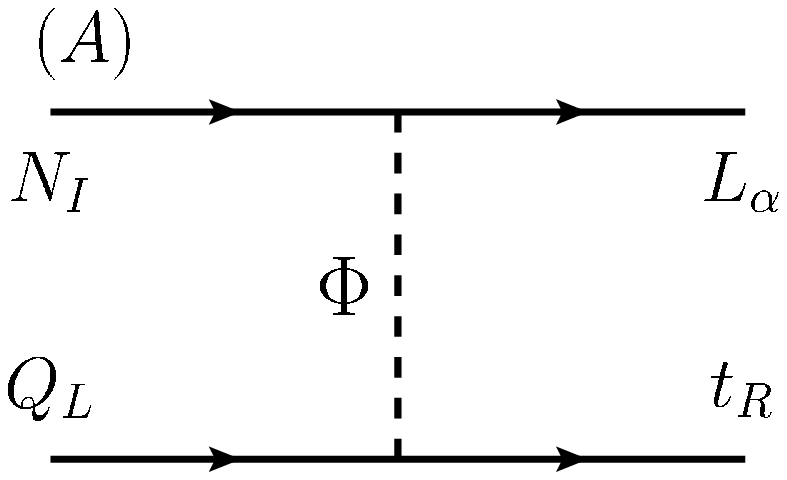}
  \includegraphics[scale=0.6]{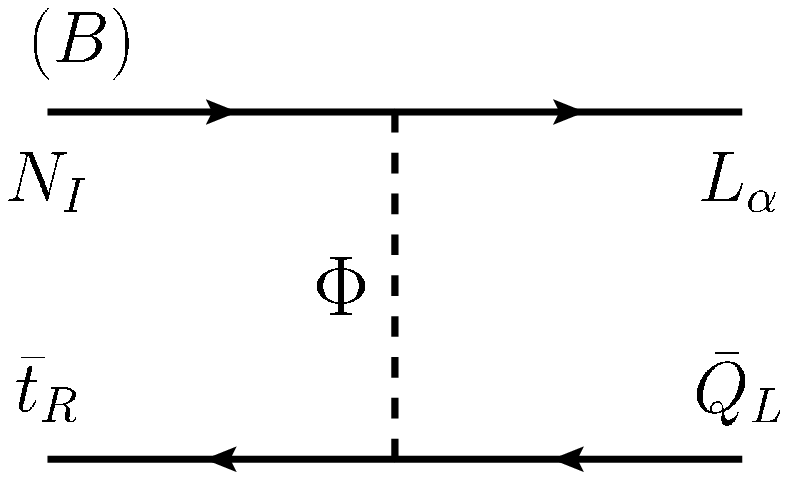}
  \includegraphics[scale=0.6]{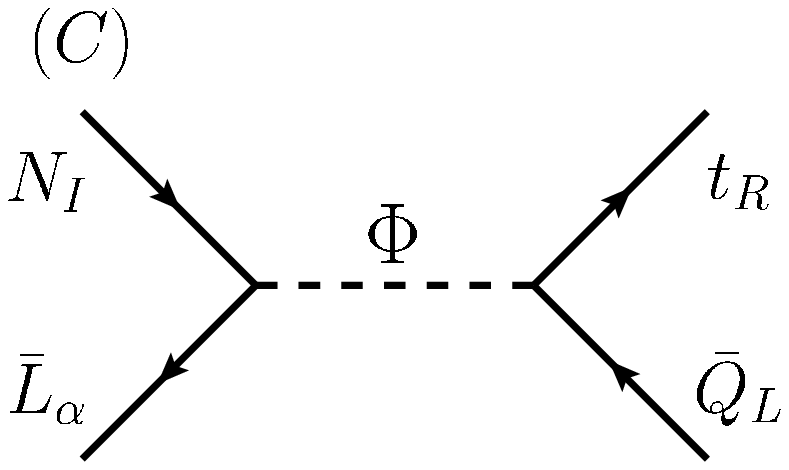}
  }%
  \caption{\it 
    Feynman diagrams for scattering processes of
    production and destruction rates.
  }
  \label{fig:Feyn}
\end{figure}
%%%%%%%%%%%%%%%%%%%%%%%%%%%%%%%%%%%%%%%%%%%%%%%%%%%%%%%%%%%%%%%%%%%%
Let us then estimate the destruction and production rates of $N_I$
with momentum $k_N$.
In the considering temperatures the dominant contributions come from
the scattering processes (A) $N_I + Q_L \leftrightarrow L_\alpha +
t_R$, (B) $N_I + \bar t_R \leftrightarrow L_\alpha + \bar Q_L$, and
(C) $N_I + \bar L_\alpha \leftrightarrow t_R + \bar
Q_L$~\cite{Akhmedov:1998qx}, shown in Fig.~\ref{fig:Feyn}. Here $Q_L$
and $t_R$ denote left-handed quark doublet of third generation and
right-handed top quark.  We then divide the rates into three parts:
\begin{eqnarray}
  \Gamma_N^{d, p} (k_N) 
  = \Gamma_N^{d,p \, {\rm (A)}} (k_N) + \Gamma_N^{d,p \, {\rm (B)}} (k_N)
  + \Gamma_N^{d,p \, {\rm (C)}} (k_N) \,.
\end{eqnarray}
The destruction rates of each process are found to be
\begin{eqnarray}
  \label{eq:GND}
  \big[ \Gamma_N^{d \, {\rm (A)}} (k_N) \big]_{IJ}
  \eqn{=} \big[ \Gamma_N^{d \, {\rm (B)}} (k_N) \big]_{IJ}
  = \gamma_N^d (k_N) \, \big[F^\dagger F \big]_{IJ} \,,
  \nonumber \\[1ex]
  \big[ \Gamma_N^{d \, {\rm (C)}} (k_N) \big]_{IJ}
  \eqn{=}
  \gamma_N^d (k_N) \, \big[ F^\dagger F \big]_{I J}
  + 
  \big[ \delta \Gamma_N^d (k_N)  \big]_{I J} \,.
\end{eqnarray}
Here we have introduced
\begin{eqnarray}
  \label{eq:gNd}
  \gamma_N^d (k_N) 
  \eqn{=} \frac{N_C N_D h_t^2}{64 \pi^3} \, \frac{T^2}{k_N} \,,
\end{eqnarray}
where $N_C=3$ is a color factor and $h_t \simeq 1$ is the top
Yukawa coupling constant, and
\begin{eqnarray}
  \big[ \delta \Gamma_N^d (k_N)  \big]_{I J}
  \eqn{=}
  \gamma_N^d (k_N) \,
  \int_0^\infty \frac{dk_L k_L}{N_D T^2}
  \Big[ F^\dagger
  \big(\rho_{\bar L}^T (k_L) - N_D \rho^{\rm eq}(k_L) \mathbf 1 \big) 
  F \Big]_{IJ} 
  \nonumber \\[1ex]
  \eqn{=} 
  \gamma_N^d (k_N) \, \big[ F^\dagger (A^{-1} - \mathbf{1}) F \big]_{IJ}  \,,
\end{eqnarray}
where we have used Eq.~(\ref{eq:RHOL}) in the last equality.
On the other hand, the production rates are
\begin{eqnarray}
  \big[ \Gamma_N^{p \, {\rm (A)}} (k_N) \big]_{IJ}
  \eqn{=}
  \big[ \Gamma_N^{p \, {\rm (B)}} (k_N) \big]_{IJ}
  =
  \gamma_N^d (k_N) \, \rho^{\rm eq}(k_N) \,
  \big[ F^\dagger F \big]_{I J}
  + \big[ \delta \Gamma_N^p (k_N) \big]_{I J} \,,
  \nonumber \\[1ex]
  \big[ \Gamma_N^{p {\rm \, (C)}} (k_N) \big]_{IJ}
  \eqn{=}
  \gamma_N^d (k_N) \, \rho^{\rm eq}(k_N) \,
  \big[ F^\dagger F \big]_{IJ} \,,
\end{eqnarray}
where 
\begin{eqnarray}
  \big[ \delta \Gamma_N^p (k_N) \big]_{I J}
  \eqn{=}
  \gamma_N^d (k_N) \, \rho^{\rm eq}(k_N) \,
  \Bigg\{
  \int_0^{k_N} \frac{d k_L}{N_D k_N}
  \frac{1 - \rho^{\rm eq}(k_L)}{\rho^{\rm eq}(k_L)}
  \Big[ F^\dagger
  \big( \rho_L (k_L) - N_D \rho^{\rm eq} (k_L) \mathbf{1} \big)
  F \Big]_{IJ}
  \nonumber \\
  \eqn{}~~~~~~~~~~~~~~~~~~~+
  \int_{k_N}^\infty \frac{d k_L}{N_D k_N}
  \frac{1 - \rho^{\rm eq}(k_N)}{\rho^{\rm eq}(k_N)}
  \Big[ F^\dagger
  \big( \rho_L (k_L) - N_D \rho^{\rm eq} (k_L) \mathbf{1} \big)
  F \Big]_{IJ} \Bigg\} 
  \nonumber \\[1ex]
  \eqn{=}
  \gamma_N^d (k_N) \, \rho^{\rm eq}(k_N) \,
  \big[ F^\dagger (A- \mathbf{1}) F \big]_{IJ} \,.
\end{eqnarray}

It should be noted that $\delta \Gamma_N^d$ and $\delta \Gamma_N^p$
vanish when all the chemical potentials of active leptons become zero
(\ie, $\rho_L (k) = \rho_{\bar L} (k) = N_D \rho^{\rm eq} (k)
\mathbf{1}$).  In this case, the rates of each process satisfy
$\Gamma_N^{p \, {\rm (A,B,C)}} (k_N) = \rho^{\rm eq}(k_N) \Gamma_N^{d \,
  {\rm (A,B,C)}} (k_N)$, and the total rates are given by
\begin{eqnarray}
  \label{eq:GNDeq}
  \big[ \Gamma_N^{d \,{\rm eq}} (k_N) \big]_{I J} 
  = 3 \, \gamma_N^d(k_N) \,
  \big[ F^\dagger F \big]_{IJ} \,,~~~
  \big[ \Gamma_N^{p \,{\rm eq}} (k_N) \big] = 
  \rho^{\rm eq}(k_N) \, \big[ \Gamma_N^{d \,{\rm eq}} (k_N) \big]_{I J} \,.
\end{eqnarray}
The kinetic equations for $\rho_N$ and $\rho_{\bar N}$ are then 
summarized as
\begin{eqnarray}
  \label{eq:KE1}
  \frac{d \rho_N(k_N)}{dt}
  \eqn{=} - i \, \big[ H_N^0(k_N) + V_N(k_N) ,\, \rho_N(k_N) \big]
  - \frac{1}{2} \, 
  \big\{ \Gamma_N^{d \, {\rm eq}} (k_N) ,\,
  \rho_N (k_N) - \rho^{\rm eq}(k_N) \mathbf{1}
  \big\}
  \nonumber \\
  \eqn{}
  + 2 \, \delta \Gamma_N^p (k_N) 
  - \frac{1}{2} \,
  \big\{ \delta \Gamma_N^d (k_N) ,\, \rho_N (k_N) \big\}
  \,.
%  \\
%  \label{eq:KE2}
%  \frac{d \rho_{\bar N}(k_N)}{dt}
%  \eqn{=} - i \, \big[ H_N^0(k_N) + V_{\bar N}(k_N) ,\, \rho_{\bar N}(k_N) \big]
%  - \frac{1}{2} \,
%  \big\{ \Gamma_{\bar N}^{d \, {\rm eq}} (k_N) ,\,
%  \rho_{\bar N} (k_N) - \rho^{\rm eq}(k_N) \mathbf{1}
%  \big\}
%  \nonumber \\
%  \eqn{}
%  + 2 \, \delta \Gamma_{\bar N}^p (k_N) 
%  - \frac{1}{2} \,
%  \big\{ \delta \Gamma_{\bar N}^d (k_N) ,\, \rho_{\bar N} (k_N) \big\}
%  \,.
\end{eqnarray}
The equation for $\rho_{\bar N}$ can be found by the CP conjugation,
\ie, by exchanging $F \leftrightarrow F^\ast$ and $\rho_L
\leftrightarrow \rho_{\bar L}$ (or $A \leftrightarrow A^{-1}$).

Next, we turn to consider the left-handed leptons.  Similar to the
above case, we start with the equation for $\rho_L$ as
\begin{eqnarray}
  \label{eq:EQRHOL}
  \frac{d \rho_L (k_L)}{dt}
  \eqn{=}
  - i \, \big[H_\nu (k_L) ,\, \rho_L (k_L) \big]
  - \frac{1}{2} \, \big\{ \Gamma_\nu^{d} (k_L) , \, \rho_L (k_L) \big\}
  + N_D \, \Gamma_\nu^{p} (k_L) \,.
%  \nonumber \\
%  \frac{d \rho_{\bar L}}{dt}
%  \eqn{=}
%  - i \, \big[H_{\bar \nu} ,\, \rho_{\bar L}  \big]
%  - \frac{1}{2} \, \big\{ \Gamma_{\bar \nu}^{d} , \, 
%    \rho_{\bar \nu} \big\}
%  + N_D \, \Gamma_{\bar \nu}^{p} \,,
\end{eqnarray}
Here remember that $\rho_L$ denotes the sum of SU(2) doublet
contributions $\rho_L = \rho_\nu + \rho_e$.  As explained in
Eq.~(\ref{eq:RHOL}), only the diagonal elements of both sides are
taken into account and then the first term of right-hand side becomes
irrelevant for the discussion.  Further, active leptons maintain the
kinetic equilibrium as Eq.~(\ref{eq:RHOL}) due to the large
interaction rates, we have only to consider the evolution of their 
chemical potentials.
\begin{eqnarray}
  \cosh \mu_{\nu_\alpha} \frac{d \mu_{\nu_\alpha}}{dt}
  = \frac{1}{4 \, N_D}
  \int \frac{dk_L k_L^2}{T^3} \,
  \frac{d}{dt} 
  \bigl[ \rho_L (k_L) - \rho_{\bar L} (k_L) \bigr]_{\alpha \alpha} \,.
\end{eqnarray}
We then estimate the destruction and production rates of active
(left-handed) neutrinos, $\Gamma_\nu^d$ and $\Gamma_\nu^p$, caused by
neutrino Yukawa interaction.  Similar to sterile neutrinos, the
dominant contributions are found from the scattering processes (A),
(B) and (C).  The production rates are then estimated as
\begin{eqnarray}
  \big[ \Gamma_\nu^{d \, {\rm (A)}} (k_L) \big]_{\alpha \beta}
  \eqn{=} \big[ \Gamma_\nu^{d \, {\rm (B)}} (k_L) \big]_{\alpha \beta}
  = \gamma_\nu^d (k_L) \, \big[ F F^\dagger \big]_{\alpha \beta} \,,
  \nonumber \\[1ex]
  \big[ \Gamma_\nu^{d \, {\rm (C)}} (k_L) \big]_{\alpha \beta}
  \eqn{=}
  \gamma_\nu^d (k_L) \big[ F F^\dagger \big]_{\alpha \beta} 
  +
  \big[ \delta \Gamma_\nu^d (k_L) \big]_{\alpha \beta} 
  \,,
\end{eqnarray}
where
\begin{eqnarray}
  \gamma_\nu^d (k_L) \eqn{=}
  \frac{N_C h_t^2 T^2}{64 \pi^3 \, k} 
  = \frac{1}{N_D} \, \gamma^d_N (k_L) \,,
  \\
  \big[ \delta \Gamma_\nu^d (k_L) \big]_{\alpha \beta} 
  \eqn{=}
  \gamma_\nu^d (k_L) 
  \int_0^\infty \frac{dk_Nk_N}{T^2} 
  \Big[ F
  \big( \rho_{\bar N}^T (k_N) - \rho^{\rm eq} (k_N) \mathbf{1} \big)
  F^\dagger \Big]_{\alpha \beta} \,.
\end{eqnarray}
On the other hand, the production rates for three processes are 
\begin{eqnarray}
  \big[ \Gamma_\nu^{p \, {\rm (A)}} (k_L) \big]_{\alpha \beta}
  \eqn{=}
  \big[ \Gamma_\nu^{p \, {\rm (B)}} (k_L) \big]_{\alpha \beta}
  =
  \gamma_\nu^d (k_L) \, \rho^{\rm eq}(k_L) \, 
  \big[ F F^\dagger \big]_{\alpha \beta}
  +
  \big[ \delta \Gamma^p_\nu (k_L) \big]_{\alpha \beta} \,,
  \nonumber \\[1ex]
  \big[ \Gamma_\nu^{p \, {\rm (C)}} (k_L) \big]_{\alpha \beta}
  \eqn{=}
  \gamma_\nu^d (k_L) \, \rho^{\rm eq}(k_L) \, 
  \big[ F F^\dagger \big]_{\alpha \beta} \,,
\end{eqnarray}
where
\begin{eqnarray}
  \big[ \delta \Gamma^p_\nu (k_L) \big]_{\alpha \beta} 
  \eqn{=}
  \gamma_\nu^d (k_L) \, \rho^{\rm eq}(k_L) \,
  \Bigg\{
    \int_0^{k_L} \frac{d k_N}{k_L}
    \frac{1 - \rho^{\rm eq}(k_N)}{\rho^{\rm eq}(k_N)}
    \Big[ F
    \big( \rho_N (k_N) - \rho^{\rm eq} (k_N) \mathbf{1} \big)
    F^\dagger \Big]_{\alpha \beta}
  \nonumber \\
  \eqn{}~~~~~~~~~~~~~~~+
    \int_{k_L}^\infty \frac{d k_N}{k_L}
    \frac{1 - \rho^{\rm eq}(k_L)}{\rho^{\rm eq}(k_L)}
    \Big[ F
    \big( \rho_N (k_N) - \rho^{\rm eq} (k_N) \mathbf{1} \big)
    F^\dagger \Big]_{\alpha \beta} \Bigg\} \,.
\end{eqnarray}
The rates of left-handed charged leptons are the same as those of
active neutrinos, and the CP conjugation ($F \leftrightarrow F^\ast$
and $\rho_N \leftrightarrow \rho_{\bar N}$) gives the rates for $\bar
\nu$.  The equations for chemical potentials are then given by
\begin{eqnarray}
  \label{eq:KE3}
  \frac{d \mu_{\nu_\alpha}}{dt} 
  \eqn{=} - 
  \gamma_\nu^d(T) \, \big[ F F^\dagger \big]_{\alpha \alpha} \, 
  \tanh \mu_{\nu_\alpha}
  \nonumber \\
  \eqn{} + \frac{\gamma^d_\nu(T)}{4} 
  \int_0^\infty \frac{dk_N k_N}{T^2}
  \Biggl\{
  \left( 1 + \frac{2}{\cosh \mu_{\nu_\alpha}} \right)
  \big[ F \rho_N(k_N) F^\dagger 
  - F^\ast \rho_{\bar N}(k_N) F^T \big]_{\alpha \alpha}
  \nonumber \\
  \eqn{} ~~~~~~~~~~~~~~~~~~~~~~~~~~~
  -
  \tanh \mu_{\nu_\alpha}
  \big[ F \rho_N(k_N) F^\dagger 
  + F^\ast \rho_{\bar N}(k_N) F^T \big]_{\alpha \alpha}
  \Biggr\} \,.
\end{eqnarray}
Therefore, the three equations, (\ref{eq:KE1}) for $\rho_N$, the CP
conjugation of (\ref{eq:KE1}) for $\rho_{\bar N}$, and (\ref{eq:KE3})
for $\mu_{\nu_\alpha}$, are the kinetic equations for our study of
baryogenesis in the $\nu$MSM.

Here are some comments.  First, we have fully taken into account the
momentum dependence in the matrices of densities $\rho_N$ and
$\rho_{\bar N}$.  As a result, the kinetic equations are written as
simultaneous integrodifferential equations, which allow us to find the
distribution of the occupation numbers of sterile neutrinos, the
significant modes to generate the baryon asymmetry, and the
temperatures crucial for baryogenesis.

Second, we have calculated the destruction and production rates of
sterile neutrinos and active leptons for a given momentum.  It should
be noted that we have taken into account the case when the initial and
final states in scattering processes which are not in the thermal
equilibrium and estimate the corrections to the rates from the
deviations $\rho_{N, \bar N} - \rho^{\rm eq}$ and $\rho_{L, \bar L} -
N_D \rho^{\rm eq}$.  Interestingly, the communication terms between
$\rho_N$ and $\rho_L$ ($\rho_{\bar N}$ and $\rho_{\bar L}$) introduced
in Ref.~\cite{Asaka:2005pn} arise automatically, which is the term
with $\delta \Gamma_N^p$ in Eq.~(\ref{eq:KE1}).  Now we can understand
well the origin of such terms.  For instance, $\delta \Gamma_N^p$
originates in the correction to the production rate of $N$ in the
scattering processes (A) and (B) when chemical potentials of active
leptons are non-zero.  The similar discussion can be applied to the
communication terms in Eq.~(\ref{eq:KE3}).  As shown in
Ref.~\cite{Asaka:2005pn}, they play a crucial role to generate the BAU
in the $\nu$MSM.

Finally, our kinetic equations also contain the communication terms
between $\rho_N$ and $\rho_{\bar L}$ ($\rho_{\bar N}$ and $\rho_{L}$)
which have not been discussed before.  For example, the term with
$\delta \Gamma^d_N$ in Eq.~(\ref{eq:KE1}) corresponds to it.  Such a
term arises as the corrections to the destruction rates of sterile
neutrinos due to the non-zero chemical potentials of active leptons
via the scattering process (C).  The important point is that the term
with $\delta \Gamma^d_N$ is the second order of the matrices of
densities (\ie, $\rho_N \rho_{\bar L}$) and the kinetic equations are
no longer linear in $\rho_N, \rho_{\bar N}$ and $\mu_{\nu_{\alpha}}$.
Further, the presence of these terms is essential to ensure the
conservation of the lepton number.

The total lepton number becomes a conserved charge for temperatures of
baryogenesis since the Majorana masses are much smaller than $T_W$, as
mentioned before.  In the considering system, this leads to
\begin{eqnarray}
  0 = \frac{d}{dt}
  \left\{ \, R^3 \,
    \left[
      \sum_{I=2,3} \left( n_{N_I} - n_{\bar N_I} \right)
      +
      \sum_{\alpha = e, \mu, \tau} \left( n_{L_\alpha} - n_{\bar L_\alpha} \right)
    \right]
  \right\} \,,
\end{eqnarray}
where $R$ is the scale factor of the expanding universe
and $n_X$ is the number density of particle $X$.
It can be written in terms of the matrices of densities
(and chemical potentials) as
\begin{eqnarray}
  \label{eq:Lconservation}
  0 \eqn{=} 
  \int \frac{d^3 k}{(2 \pi)^3} \,
  \frac{d}{dt} \,
  \mbox{tr}
  \bigl[ \rho_N(k) - \rho_{\bar N}(k) 
    + \rho_L (k) - \rho_{\bar L}(k)
  \bigr]
  \nonumber \\
  \eqn{=}
  \int \frac{d^3 k}{(2 \pi)^3} \,
  \frac{d}{dt} \,
  \mbox{tr} \bigl[ \rho_N(k) - \rho_{\bar N}(k) \bigr]
  + \frac{2 N_D T^3}{\pi^2} \cosh \mu_{\nu_\alpha} 
  \frac{d \mu_{\nu_\alpha}}{dt} \,.
\end{eqnarray}
A simple calculation shows that our equations (\ref{eq:KE1}) and
(\ref{eq:KE3}) satisfy this equality.
It is found for the scattering processes (A) and (B) that the
production (or destruction) terms in $\rho_N$ and $\rho_{\bar N}$
cancel with the destruction (or production) terms in $\rho_L$ and
$\rho_{\bar L}$, respectively, after taking the trace and integrating
over momentum.  As for the process (C) the production (or destruction)
terms in $\rho_N$ and $\rho_{\bar N}$ cancel with the production (or
destruction) terms in $\rho_{\bar L}$ and $\rho_{L}$, respectively.
Thus, if the system starts with the lepton symmetric universe, the
asymmetry of sterile neutrinos is always opposite to that of
active leptons.  Namely, the kinetic equations here describe not the
generation of the total lepton asymmetry (\ie, {\it leptogenesis}), but
the separation into sterile and active sectors.

%%%%%%%%%%%%%%%%%%%%%%%%%%%%%%%%%%%%%%%%%%%%%%%%%%%%%%%%%%%%%%%%%%%%
%%%%%%%%%%%%%%%%%%%%%%%%%%%%%%%%%%%%%%%%%%%%%%%%%%%%%%%%%%%%%%%%%%%%
\section{Numerical Solution of Kinetic Equations}
\label{sec:NS}
%%%%%%%%%%%%%%%%%%%%%%%%%%%%%%%%%%%%%%%%%%%%%%%%%%%%%%%%%%%%%%%%%%%%
%%%%%%%%%%%%%%%%%%%%%%%%%%%%%%%%%%%%%%%%%%%%%%%%%%%%%%%%%%%%%%%%%%%%
In this section, we shall study the numerical solution of the kinetic
equations (\ref{eq:KE1}) and (\ref{eq:KE3}).  To incorporate the
expansion of the universe we replace the time derivative in
Eq.~(\ref{eq:KE1}) as $\frac{d}{dt} \to \frac{\partial}{\partial t} -
H k \frac{\partial}{\partial k}$, where $H = T^2/M_0$ ($M_0 = 7.12
\times 10^{17}$ GeV) is the Hubble expansion rate.  The
time-temperature relation is then given by $ \frac{dT}{dt} =
\frac{T^3}{M_0}$.  In this analysis, we take the initial conditions as
$\rho_N (k)=\rho_{\bar N}(k) = \mu_{\nu_\alpha}=0$, which may be
realized by the primordial inflation.  We solve these equations till
$T=T_W$ in order to estimate the BAU, where $T_W$ is the sphaleron
freezing temperature. (See, for example Ref.~\cite{hep-ph/0511246}.)
Following to Ref.~\cite{Canetti:2010aw}, we set $T_W=140$ GeV from
now on.

In solving the kinetic equations, we have to specify neutrino Yukawa
coupling constants $F_{\alpha I}$ and masses of sterile neutrinos
$M_2$ and $M_3$.  The estimation of the baryon asymmetry in full
parameter region is beyond the scope of this analysis, which is
postponed in another publication.  Here we choose a specific choice
of parameters and study the properties of the solutions.

As for the neutrino Yukawa couplings we follow the notation in
Ref.~\cite{Asaka:2011pb}.  In this analysis we consider the normal
hierarchy of active neutrino masses and take $m_3 = 4.89 \times
10^{-2}$ eV, $m_2 = 8.71 \times 10^{-3}$ eV, $m_1 =0$, 
$\sin^2 \theta_{23}=0.42$, $\sin^2
\theta_{12}=0.312$ and $\sin^2 \theta_{13}
= 0.025$~\cite{arXiv:1106.6028}.  
Further, we choose the parameters of
sterile neutrinos as $\xi = +1$, $\omega = \pi/4$,
$\delta = 7 \pi/4$ and $\eta = \pi/3$.  We
write masses of sterile neutrinos  as $M_3 = M_N + \Delta M_N/2$
and $M_2 = M_N -\Delta M_N/2$, and choose, as representative values,
 $M_N = 10$ GeV by fixing $\Delta M_N/M_N = 10^{-8}$ otherwise stated.

%%%%%%%%%%%%%%%%%%%%%%%%%%%%%%%%%%%%%%%%%%%%%%%%%%%%%%%%%%%%%%%%%%%%
%%%%% ** Figure ** %%%%%%%%%%%%%%%%%%%%%%%%%%%%%%%%%%%%%%%%%%%%%%%%%
\begin{figure}[t]
  \centerline{
  \includegraphics[scale=1.5]{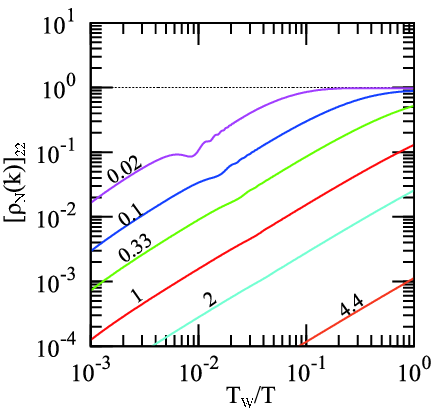}%
  \hspace{1cm}
  \includegraphics[scale=1.5]{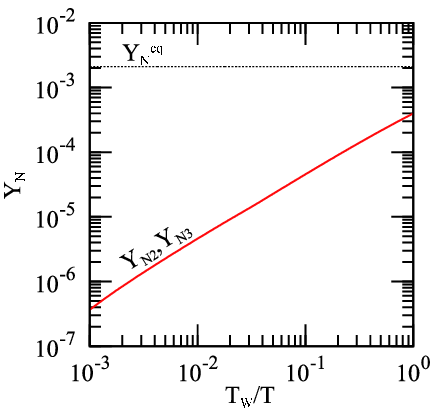}%
  }%
  \caption{\it 
    Evolution of $[\rho_N(k)]_{22}$ with $k/T=$ 0.02,
    0.1, 0.33, 1, 2 and 4.4 in the left panel,
    and evolution of $Y_{N_{2,3}}$ in the right panel.
  }
  \label{fig:EVO_YN}
\end{figure}
%%%%%%%%%%%%%%%%%%%%%%%%%%%%%%%%%%%%%%%%%%%%%%%%%%%%%%%%%%%%%%%%%%%%
The diagonal elements of the matrices of densities $\rho_N$ and
$\rho_{\bar N}$ are the occupation numbers of $N_I$ and $\bar N_I$.
The evolution of these quantities are shown in the left panel of
Fig.~\ref{fig:EVO_YN}.  In the present choice of parameters (\ie,
$\omega = \pi/4$) the occupation numbers of $N_2$ and $N_3$ are almost
the same, and we only show the evolution of $[\rho_N]_{22}$.  It is
seen that the occupation numbers increase as the temperature
decreases due to the production by the scattering processes (A), (B)
and (C).  In Eq.~(\ref{eq:KE1}) the third term on the right-hand side
which proportional to $\rho^{\rm eq}$ is the main source of the
production of sterile neutrinos.  Since the destruction rate in
Eq.~(\ref{eq:GNDeq}) is inversely proportional to the momentum, the
production of modes with lower momenta is more effective.  We find
that the low modes stop to grow eventually and take constant values
given by $[\rho_N(k)]_{II}=[\rho_{\bar N}(k)]_{II} = \rho^{\rm eq}(k)$
afterward.  This thermalization temperature becomes higher
for the mode with smaller momentum.

The yields of $N_I$ and $\bar N_I$, defined by the ratio between
the number and entropy densities, are given by
\begin{eqnarray}
  Y_{N_I} = \frac{1}{s}
  \int \frac{d^3 k}{(2\pi)^3} \,
  \big[ \rho_N (k) \big]_{II} \,,~~~
  Y_{\bar N_I} = \frac{1}{s}
  \int \frac{d^3 k}{(2\pi)^3} \,
  \big[ \rho_{\bar N} (k) \big]_{II} \,,
\end{eqnarray}
where the entropy density is $s = \frac{2 \pi^2}{45} g_s T^3$ 
with $g_s = 106.75$.  When the system is fully thermalized
(\ie, $[\rho_N (k)] = [\rho_{\bar N}(k)] = \rho^{\rm eq}(k) \mathbf{1}$),
the yields take a constant value 
\begin{eqnarray}
  Y_{N_I}^{\rm eq} = Y_{\bar N_I}^{\rm eq} = 
  \frac{45}{2 g_s \pi^4} = 2.1 \times 10^{-3} \,,
\end{eqnarray}
for the temperatures of interest.
It is found that $Y_{N_2}$ and $Y_{N_3}$ increase as $1/T$
until they close to the equilibrium values
as shown in Fig.~\ref{fig:EVO_YN}.

%%%%%%%%%%%%%%%%%%%%%%%%%%%%%%%%%%%%%%%%%%%%%%%%%%%%%%%%%%%%%%%%%%%%
%%%%% ** Figure ** %%%%%%%%%%%%%%%%%%%%%%%%%%%%%%%%%%%%%%%%%%%%%%%%%
\begin{figure}[t]
  \centerline{
    \includegraphics[scale=1.4]{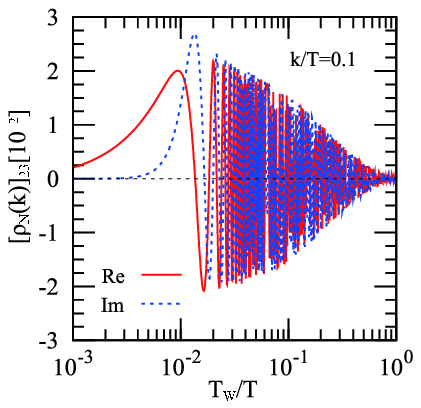}%
    \includegraphics[scale=1.4]{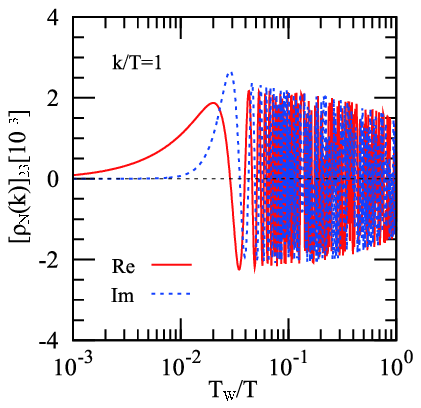}%
    \includegraphics[scale=1.4]{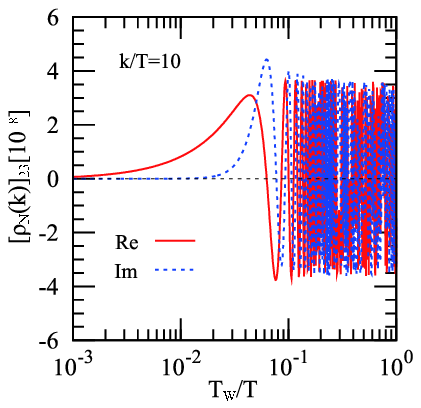}%
  }%
  \caption{\it 
    Evolution of $\mbox{\rm Re}[\rho_N]_{23}$ (red solid lines) and 
    $\mbox{\rm Im}[\rho_{N}]_{23}$ (blue dashed lines) 
    for the modes with $k/T =$ 0.1 (left panel), 1 (middle panel)
    and 10 (right panel), respectively.
  }
  \label{fig:EVO_N23}
\end{figure}
%%%%%%%%%%%%%%%%%%%%%%%%%%%%%%%%%%%%%%%%%%%%%%%%%%%%%%%%%%%%%%%%%%%%
On the other hand, the off-diagonal elements of the matrices of
densities contain correlations of the flavor mixing.  In
Fig.~\ref{fig:EVO_N23} we show the evolution of $[\rho_N(k)]_{23}$ for
the modes with $k/T=0.1$, 1 and 10.  It is seen that both real and
imaginary parts of $[\rho_N(k)]_{23}$ start to oscillate around the
temperature~\cite{Asaka:2005pn}
\begin{eqnarray}
  T_{\rm osc}(k) = \left( \frac{M_0 \, \Delta M_N \, M_N}
    {3 \, (k/T)} \right)^{1/3}  \,,
\end{eqnarray}
and hence the oscillation for the mode with lower momentum begins at
higher temperature.  We should note that the amplitude of the
oscillation becomes damped and the off-diagonal elements of $\rho_N$
vanish when the mode gets in the thermal equilibrium.  This behavior
can be seen for the mode with $k/T=0.1$ in Fig.~\ref{fig:EVO_N23}.

The flavor oscillation between $N_2$ and $N_3$ can generate the
asymmetries of sterile neutrinos as well as active
leptons~\cite{Akhmedov:1998qx,Asaka:2005pn}, together with the CP
violation in neutrino sector.  The analytical description of these
production processes are found in
Refs.~\cite{Akhmedov:1998qx,Asaka:2005pn,Shaposhnikov:2008pf,Asaka:2010kk},
and we just show the results of the numerical analysis.

%%%%%%%%%%%%%%%%%%%%%%%%%%%%%%%%%%%%%%%%%%%%%%%%%%%%%%%%%%%%%%%%%%%%
%%%%% ** Figure ** %%%%%%%%%%%%%%%%%%%%%%%%%%%%%%%%%%%%%%%%%%%%%%%%%
\begin{figure}[t]
  \centerline{
  \includegraphics[scale=1.5]{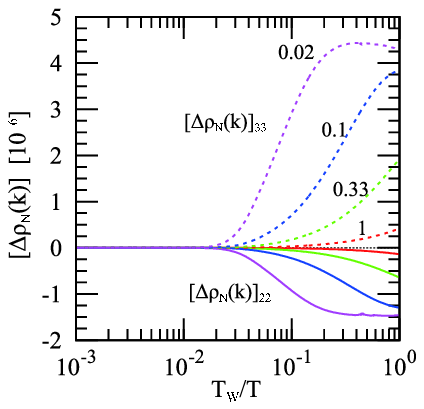}%
  \hspace{1cm}
  \includegraphics[scale=1.5]{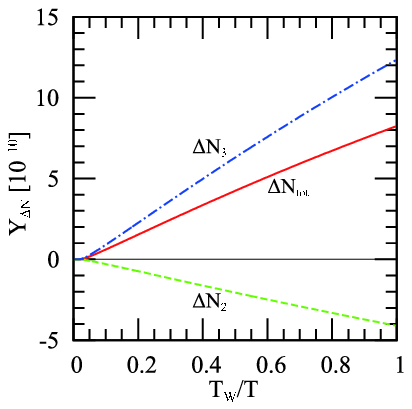}%
  }%
  \caption{\it 
    Evolution of $[\Delta \rho_N]_{22}$ (solid lines)
    and $[\Delta \rho_N]_{33}$ (dotted lines)
    with momenta $k/T =$ 0.02, 0.1, 0.33 and 1
    in the left panel,
    and evolution of $Y_{\Delta N_2}$, $Y_{\Delta N_3}$ and 
    $Y_{\Delta N_{\rm tot}}$ in the right panel.
  }
  \label{fig:EVO_YDN}
\end{figure}
%%%%%%%%%%%%%%%%%%%%%%%%%%%%%%%%%%%%%%%%%%%%%%%%%%%%%%%%%%%%%%%%%%%%
The asymmetries in the occupation numbers of sterile neutrinos,
$[\Delta \rho_N(k)]_{II} = [\rho_N(k)]_{II}-[\rho_{\bar N}(k)]_{II}$,
are generated as shown in the left panel of Fig.~\ref{fig:EVO_YDN}.
It is seen that such an asymmetry for the mode with $k$ is generated
at $T \simeq T_{\rm osc}(k)$ and the absolute value of the asymmetry
becomes larger for the lower modes.  We also find that the asymmetries
of the mode with $k/T = 0.02$ begins to damp at $T_W/T \simeq 0.2$, at
which the mode gets in thermal equilibrium.  Having the asymmetries of
each mode, we can estimate the asymmetry in the number density as
\begin{eqnarray}
  Y_{\Delta N_I} = \frac{1}{s}\int \frac{d^3k}{(2\pi)^3}
  \big[ \Delta \rho_N (k) \big]_{II} \,.
\end{eqnarray}
As shown in the right panel of Fig.~\ref{fig:EVO_YDN}, $Y_{\Delta
  N_I}$ increases as $\sim 1/T$ as long as the system is away from the
equilibrium state.  We also show the evolution of total asymmetry of
sterile neutrinos, $Y_{\Delta N_{\rm tot}} = Y_{\Delta N_2} +
Y_{\Delta N_3}$.

%%%%%%%%%%%%%%%%%%%%%%%%%%%%%%%%%%%%%%%%%%%%%%%%%%%%%%%%%%%%%%%%%%%%
%%%%% ** Figure ** %%%%%%%%%%%%%%%%%%%%%%%%%%%%%%%%%%%%%%%%%%%%%%%%%
\begin{figure}[tb]
  \centerline{
  \includegraphics[scale=1.5]{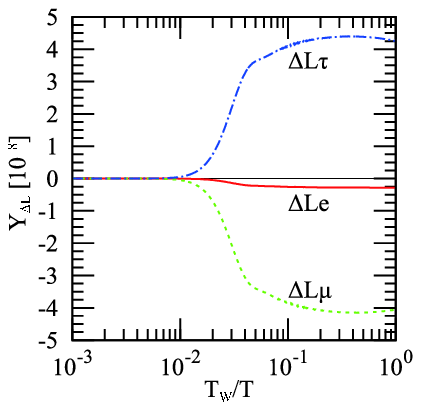}%
  \hspace{1cm}
  \includegraphics[scale=1.5]{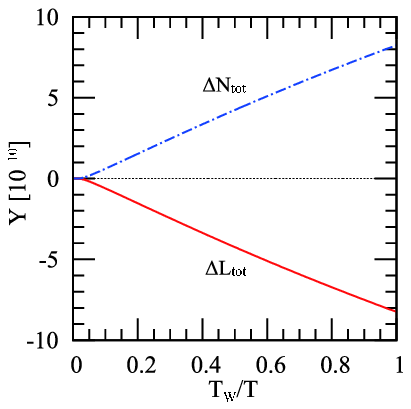}%
  }%
  \caption{\it Evolution of $Y_{\Delta L_e}$, $Y_{\Delta L_\mu}$ 
    and $Y_{\Delta L_\tau}$ in the left panel, and evolution of 
    $Y_{\Delta L_{\rm tot}}$ and $Y_{\Delta N _{\rm tot}}$ 
    in the right panel.  }
  \label{fig:EVO_YDL}
\end{figure}
%%%%%%%%%%%%%%%%%%%%%%%%%%%%%%%%%%%%%%%%%%%%%%%%%%%%%%%%%%%%%%%%%%%%
In the considering system, the non-zero asymmetry in the sterile
sector necessarily indicates the presence of asymmetry in active
sector.  The yields of active flavor asymmetries are expressed in
terms of chemical potentials as
\begin{eqnarray}
  Y_{\Delta L_\alpha} = \frac{45 N_D}{g_s \pi^4} \sinh \mu_\alpha \,.
\end{eqnarray}
The evolution of $Y_{\Delta L_\alpha}$ is shown in the left panel of
Fig.~\ref{fig:EVO_YDL}.  The asymmetries of active flavors $Y_{\Delta
  L_\alpha}$ are larger than those of sterile neutrinos $Y_{\Delta
  N_I}$.  This is because $Y_{\Delta L_\alpha}$ is induced at the
fourth order of neutrino Yukawa coupling constants, while $Y_{\Delta
  N_I}$ is at the sixth order~\cite{Asaka:2005pn}.  As shown in
Ref.~\cite{Asaka:2010kk} when $\mbox{Im}\omega =0$, the yield
$Y_{\Delta L_\alpha}$ is proportional to the CP asymmetry parameter
$A_{32}^\alpha$.  In the present choice of parameters, they are
\begin{eqnarray}
  A_{32}^e : A_{32}^\mu : A_{32}^\tau =
  - 1.0 : -15 : 16 \,.
\end{eqnarray}
It is then found that our numerical solutions of the kinetic equations
reproduce the analytic relations of asymmetries of active lepton
flavors.

The total asymmetry stored in active leptons is now obtained as
$Y_{\Delta L_{\rm tot}} = Y_{\Delta L_e} + Y_{\Delta L_\mu} +
Y_{\Delta L_\tau}$, which evolution is shown in the right panel of
Fig.~\ref{fig:EVO_YDL}.  For comparison we also show the asymmetry in
sterile sector.  It is clearly seen that $Y_{\Delta N_{\rm tot}} +
Y_{\Delta L_{\rm tot}} = 0$ due to the conservation of the total
lepton number in the considering system.  These total asymmetries are
obtained as the ${\cal O}(F^6)$ effects.  We also find that the
generation of the total asymmetries becomes effective at the
temperature
\begin{eqnarray}
  T_L \sim T_{\rm osc} (k = T) \,.
\end{eqnarray}
We find $T_L \sim 6.2 \times 10^3$ GeV for our choice of parameters.
These yields scale as $1/T$, which are consistent with the estimation
in Ref.~\cite{Asaka:2005pn}, even if one includes the momentum
dependence of the matrices of densities properly.

The asymmetry in active sector can be partially converted into 
the baryon asymmetry as $\Delta B = - \frac{28}{79} \Delta L_{\rm tot}$
due to the rapid sphaleron transition~\cite{Kuzmin:1985mm}.
The BAU observed in the present universe is then given by
\begin{eqnarray}
  Y_B = - \frac{28}{79} \left. Y_{\Delta L_{\rm tot}} \right|_{T=T_W} 
  = - 1.53 \times 10^{-3} N_D \sum_\alpha  \left. 
    \sinh \mu_{\nu_\alpha}
    \right|_{T=T_W} \,.
\end{eqnarray}
Notice that the asymmetry for $T < T_W$ cannot contribute
to the BAU since the sphaleron process is ineffective for the lower
temperatures.  In the considering parameter choice,
the BAU is given by 
(See also the case III in Tab.~\ref{tab:YB}.)
\begin{eqnarray}
  Y_B = \left\{
    \begin{array}{l l}
      2.9 \times 10^{-10} & ~~\mbox{for}~ M_N = 10 \GeV 
      \\[1ex]
      5.4 \times 10^{-14} & ~~\mbox{for}~ M_N = 100 \MeV 
    \end{array}
    \right. \,.
\end{eqnarray}

%%%%%%%%%%%%%%%%%%%%%%%%%%%%%%%%%%%%%%%%%%%%%%%%%%%%%%%%%%%%%%%%%%%%
%%%%% ** Figure ** %%%%%%%%%%%%%%%%%%%%%%%%%%%%%%%%%%%%%%%%%%%%%%%%%
\begin{figure}[t]
  \centerline{
  \includegraphics[scale=1.5]{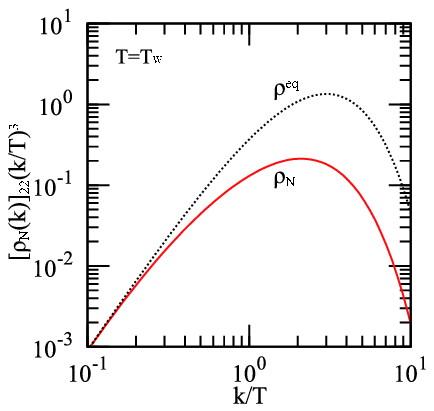}%
  \hspace{1cm}
  \includegraphics[scale=1.5]{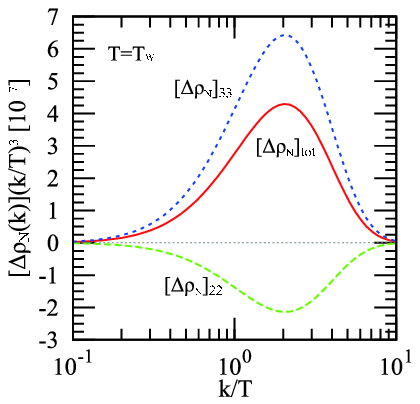}%
  }%
  \caption{\it 
    Momentum distributions of $[\rho_N(k)]_{22}$ (red solid line) and
    $\rho^{\rm eq}(k)$ (black dotted line) in the left panel.
    Momentum distributions of $[\Delta \rho_N(k)]_{22}$ 
    (green dashed line), $[\Delta \rho_N(k)]_{33}$
    (blue dotted line), and $[\Delta \rho_N(k)]_{\rm tot}$ (red
    solid line) in the right panel. The
    distributions are evaluated at $T=T_W$ and are multiplied by
    $(k/T)^3$.  }
  \label{fig:SPECN22}
\end{figure}
%%%%%%%%%%%%%%%%%%%%%%%%%%%%%%%%%%%%%%%%%%%%%%%%%%%%%%%%%%%%%%%%%%%%
When the sphaleron process is frozen at $T=T_W$, the distribution of
the occupation number $[\rho_N (k)]_{22} \simeq [\rho_N (k)]_{33}$ is
represented in terms of momentum in the left panel of
Fig.~\ref{fig:SPECN22}.  We should stress here that the momentum
dependence in the occupation number $[\rho_N(k)]_{II}$ is
significantly different from the equilibrium one $\rho^{\rm eq}(k)$.
This is because of the momentum dependence in the destruction and
production rates of sterile neutrinos estimated in the previous
section.  These rates are characterized by $\gamma_N^d(k)$ in
Eq.~(\ref{eq:gNd}) and $\gamma_N^d(k) \propto 1/k$.  This results in
the enhancement and the suppression of the production of the mode with
lower and higher momentum than about $T$, respectively.  Especially,
the low modes with $k/T \lesssim 0.1$ are equilibrated until $T =
T_W$ and coincides with $\rho^{\rm eq}(k)$.  It is also found that the
most significant mode for the number density is found to be $k \simeq
2T$, which is smaller than $\langle k \rangle = 3T$ obtained by the
thermal average.

In the right panel of Fig.~\ref{fig:SPECN22} we show the momentum
distributions of the asymmetries of the occupation numbers, $[\Delta
\rho_N(k)]_{22}$, $[\Delta \rho_N(k)]_{33}$ and their sum $[\Delta
\rho_N(k)]_{\rm tot}$, at $T=T_W$.  Similar to the occupation numbers,
they have different momentum distributions from $\rho^{\rm eq}(k)$.
(See Fig.~\ref{fig:COMP}.)  Interestingly, the most important mode for
the total asymmetry in sterile sector (and hence for the BAU) is found
to be $k \simeq 2 T$ as before.

Therefore, these characteristic behaviors of the distributions show
that the momentum dependence in the matrices of densities (as well as
the destruction and production rates) is indispensable for the correct
evaluation of the BAU in the $\nu$MSM.
%%%%%%%%%%%%%%%%%%%%%%%%%%%%%%%%%%%%%%%%%%%%%%%%%%%%%%%%%%%%%%%%%%%%
%%%%%%%%%%%%%%%%%%%%%%%%%%%%%%%%%%%%%%%%%%%%%%%%%%%%%%%%%%%%%%%%%%%%
\section{Comparison with previous works}
\label{sec:Comparison}
%%%%%%%%%%%%%%%%%%%%%%%%%%%%%%%%%%%%%%%%%%%%%%%%%%%%%%%%%%%%%%%%%%%%
%%%%%%%%%%%%%%%%%%%%%%%%%%%%%%%%%%%%%%%%%%%%%%%%%%%%%%%%%%%%%%%%%%%%
In this section we shall compare the results obtained in this analysis
with the previous works, especially given in Ref.~\cite{Asaka:2010kk},
and also estimate the qualitative difference on the BAU.
So far the momentum dependence of $\rho_N$ and $\rho_{\bar N}$ have
been approximately taken into account.  
To make the differences clear, 
let us first consider what kind of approximations should be imposed on
the kinetic equations~(\ref{eq:KE1}) and (\ref{eq:KE3}) 
in order to reproduce the previous results.

It is then found that there are three important assumptions.
(i) The momentum dependence 
is approximated by the equilibrium one, namely, 
\begin{eqnarray}
  \label{eq:AP_RHON}
  \big[ \rho_N (k) \big]_{IJ} = \big[ R_N \big]_{IJ} \, \rho^{\rm eq}(k) \,,
  ~~~
  \big[ \rho_{\bar N} (k) \big]_{IJ} = \big[ R_{\bar N} \big]_{IJ} \, 
  \rho^{\rm eq}(k) \,,
\end{eqnarray}
and the evolution of $R_N$ and $R_{\bar N}$ is studied.  (ii) The
thermal average is taken for the destruction rates of sterile neutrinos.
(iii) The destruction and production rates by the scattering
process (C) are identified with those by the process (A) or (B).
We will discuss these points in turn.

First of all, based on the assumption (i),
the equations for $R_N$ and chemical potentials are 
from Eqs. (\ref{eq:KE1}) and (\ref{eq:KE3}) as
\begin{eqnarray}
  \label{eq:KEAP_RN1}
  \frac{d R_N}{dt }
  \eqn{=} - i \big[ H_N(k), \, R_N \big] 
  - \frac{3 \, \gamma_N^d (k)}{2}  \, 
  \big\{ F^\dagger F, R_N - \mathbf{1} \big\}
  + 2 \, \gamma_N^d (k) \ F^\dagger ( A - \mathbf{1} ) F
  \nonumber \\
  \eqn{}
  - \frac{\gamma_N^d (k)}{2}  \, 
  \big\{ F^\dagger (A^{-1} - \mathbf{1} ) F, R_N \big\} \,,
  \\[1ex]
  \label{eq:KEAP_MU}
  \frac{d \mu_{\nu_\alpha}}{dt}
  \eqn{=}
  - \frac{3 \, \gamma_{\nu}^d (T)}{2}  \, \big[ F F^\dagger \big]_{\alpha \alpha}
  \tanh \mu_\alpha
  + \frac{\gamma_\nu^d (T)}{2} 
  \big[ F R_N F^\dagger - F^\ast R_{\bar N} F^T \big]_{\alpha \alpha}
  \frac{1}{\cosh \mu_\alpha}
  \nonumber \\
  \eqn{}
  + \frac{\gamma_\nu^d(T)}{4} 
  \Bigl\{
  \big[ F (R_N- \mathbf{1}) F^\dagger \big]_{\alpha \alpha}  (1 - \tanh \mu_\alpha)
  - \big[ F^\ast (R_{\bar N} - \mathbf{1})F^T \big]_{\alpha \alpha}
    ( 1 + \tanh \mu_\alpha ) \Bigr\} \,.
\end{eqnarray}
The equation for $R_{\bar N}$ can be found as explained in
Sec.~\ref{sec:KE}.  We then obtain the coupled differential equations
rather than the coupled integrodifferential equations.
Let us recall here the origins of the communication terms 
in these equations.
As for Eq.~(\ref{eq:KEAP_RN1})
the third term comes from $\delta \Gamma_N^p$ by the processes (A)
and (B) and the fourth term comes from $\delta \Gamma_N^d$
by the process (C).  Similarly,
$\delta \Gamma_\nu^p$ by (A) and (B) induces the third term
and $\delta \Gamma_\nu^d$ by (C) gives the fourth term
in Eq.~(\ref{eq:KEAP_MU}).

We can see that, although $R_N$ is introduced to be independent on the
momentum $k$, the equation (\ref{eq:KEAP_RN1}) does depend on $k$
through $\gamma_N^d(k)$ to which all the destruction and production
rates of sterile neutrinos are proportional, and also through the
effective Hamiltonian $H_N(k)$.  Note that the rates of chemical
potentials are independent on the momentum since they are proportional
to $\gamma_\nu^d(T)$.  In order to avoid this difficulty one might
consider that the momentum in Eq.~(\ref{eq:KEAP_RN1}) is replaced as
$k = k_\ast \sim T$, since such a mode is expected to
give the dominant contribution to the number densities and also the
asymmetries.  However, we have to carefully
choose $k_\ast$ to keep the conservation of the total lepton number.
Under the considering situation, Eq.~(\ref{eq:Lconservation}) is
satisfied when
\begin{eqnarray}
  \label{eq:LC2}
  0 \eqn{=}
  \mbox{tr}
  \left[
   \left. \frac{d R_N}{dt} \right|_{k = k_\ast}
   -
   \left. \frac{d R_{\bar N}}{dt} \right|_{k = k_\ast}
   + N_D \frac{dA}{dt} - N_D \frac{d A^{-1} }{dt}
  \right] \,.
\end{eqnarray}
It can be shown that it is fulfilled if and
only if $k_\ast = 2 T$.  

Interestingly, we observe that 
the rates with $k = 2 T$ correspond to nothing
but those obtained by the thermal average.  To see this point, let
us introduce the averaged $\gamma_N^d$ by
\begin{eqnarray}
  \langle \gamma^d_N \rangle
  \equiv
  \frac{\displaystyle
    \int \frac{d^3k}{(2\pi)^3} \, \rho^{\rm eq}(k) \, \gamma_N^d (k) }
  {\displaystyle
    \int \frac{d^3k}{(2\pi)^3} \, \rho^{\rm eq}(k) }
  = \frac{N_C N_D h_t^2 T}{128 \pi^3} 
  \,.
\end{eqnarray}
and hence $\langle \gamma^d_N \rangle = \left. \gamma^d_N(k) \right|_{k
  = 2T} $.  Note that $\gamma_\nu^d(T) = \langle \gamma^d_N  \rangle$,
which is accidentally obtained because of $N_D = 2$.
Therefore, the assumption (ii), \ie,
the thermal average of the destruction rates in
Eq.~(\ref{eq:GND}), can then be achieved by the replacement $\gamma_N^d
(k) \to \langle \gamma_N^d \rangle$.
It should be noted that the averaged $\Gamma_N^{d \, {\rm eq}}(k)$
in Eq.~(\ref{eq:GNDeq}) is given by
\begin{eqnarray}
  \langle \Gamma_N^{d \, {\rm eq}} \rangle
  = 3 \langle \gamma_N^d \rangle \, \big[ F^\dagger F \big]
  = \frac{9 h_t^2}{64 \pi^3} \, T \, \big[ F^\dagger F \big] \,,
\end{eqnarray}
which is exactly the (total) destruction rate of sterile neutrinos
introduced by Ref.~\cite{Akhmedov:1998qx} and used in 
Refs.~\cite{Asaka:2005pn,Asaka:2010kk,Canetti:2010aw}.

In the end, by applying the assumptions (i) and (ii),
our kinetic equations lead to the equation for $R_N$ 
\begin{eqnarray}
  \label{eq:KEAP_RN}
  \frac{d R_N}{dt }
  \eqn{=} - i \big[ \langle H_N \rangle, \, R_N \big] 
  - \frac{3}{2} \langle \gamma_N^d \rangle \, 
  \big\{ F^\dagger F, R_N - \mathbf{1} \big\}
  + 2 \, \langle \gamma_N^d \rangle \ F^\dagger ( A - \mathbf{1} ) F
  \nonumber \\
  \eqn{}
  - \frac{1}{2} \, \langle \gamma_N^d \rangle \, 
  \big\{ F^\dagger (A^{-1} - \mathbf{1} ) F, R_N \big\} \,.
\end{eqnarray}
We should compare the equations (\ref{eq:KEAP_RN})
and (\ref{eq:KEAP_MU})  with those in the previous works.
Notice that they satisfy Eq.~(\ref{eq:LC2})
and hence the lepton number is conserved for each mode
with~$k$.

In Ref.~\cite{Akhmedov:1998qx} the evolution of $R_N$ and $R_{\bar N}$
has been studied by using Eq.~(\ref{eq:KEAP_RN}) where only the first
two terms are taken into account and include no effect of chemical
potentials.  Ref.~\cite{Asaka:2005pn} has solved the coupled
equations of $R_{N, \bar N}$ and chemical potentials (the equations
for $\rho_{L, \bar L}$ in~\cite{Asaka:2005pn} are nothing but those of
chemical potentials), where the third term in (\ref{eq:KEAP_RN}) and
the first two terms in (\ref{eq:KEAP_MU}) are taken into account,
however coefficients are different from the present ones.  See the
discussion below.  The last terms in Eqs.~(\ref{eq:KEAP_RN}) and
(\ref{eq:KEAP_MU}) are original ones in the present paper.

We shall discuss in detail the differences from
Ref.~\cite{Asaka:2010kk} where SU(2) degrees of freedom of lepton
doublets are taken into account.%
\footnote{ The kinetic equations in~\cite{Asaka:2010kk} are reproduced
  from those in~\cite{Asaka:2005pn,Canetti:2010aw} by replacing
  $\Gamma_L \to \Gamma_L/N_D$ and $\mu_{\alpha} \to N_D \mu_{\alpha}$.
  The change of $\Gamma_L$ leads to the change of coefficient of the
  communication term in the equation of $\rho_N$ accordingly.  The
  final value of the BAU in~\cite{Asaka:2010kk} decreases roughly by a
  factor of two due to these modifications.  } 
As for the equation (\ref{eq:KEAP_RN}) there are two differences.  One
is the coefficient of the third term is ``2'' rather than ``3''.  The
other one is the last term is missing in Ref.~\cite{Asaka:2010kk}.
(Correspondingly, the equations for chemical potentials are also
different.)  We observe that the equations in Ref.~\cite{Asaka:2010kk}
can be reproduced if one uses the destruction and production rates
from the scattering process (A) or (B) (the rates for these two
processes are the same as shown in Sec.~\ref{sec:KE}), and multiplies
the rates a factor of three to get the total rates.  This is the
assumption (iii) listed before.

As explained in Sec.~\ref{sec:KE}, the processes (A) and (B) lead to
the communication terms between $\rho_N$ and $\rho_L$, which induce
the third term of Eq.~(\ref{eq:KEAP_RN}).  On the other hand, the
process (C) connects between $\rho_N$ and $\rho_{\bar L}$ and gives
the fourth term of Eq.~(\ref{eq:KEAP_RN}).  Note again that these
communication terms arise as the corrections to the destruction and
production rates of particles, not by hand.  As a result, we can find
the explicit form of such a term with a definite coefficient.

%%%%%%%%%%%%%%%%%%%%%%%%%%%%%%%%%%%%%%%%%%%%%%%%%%%%%%%%%%%%%%%%%%%%
\renewcommand{\arraystretch}{1.5}
\begin{table}
  \centering
  \begin{tabular}[t]{| c || c | c || c |}
    \hline
    \makebox[80pt][c]{Case} & \makebox[4cm][c]{I} &
    \makebox[4cm][c]{II} & \makebox[4cm][c]{III}\\
%    \hline 
%    Case & I & II & III 
%    \\
    \hline \hline
    {Kinetic eqs.} & 
    {Ref.~\cite{Asaka:2010kk}} & 
    Eqs.~(\ref{eq:KEAP_MU}) and (\ref{eq:KEAP_RN}) &
    Eqs. (\ref{eq:KE1}) and (\ref{eq:KE3})
    \\ \hline
    Momentum dep. &
    Approx. (i), (ii), (iii) &
    Approx. (i), (ii) &
%    \multicolumn{2}{c ||}{Approximations (i) and (ii)}  &
    Full
    \\ \hline
    Rates & 
    $\Gamma^d_N = \langle 3 \, \Gamma^{d \, {\rm (A)}} \rangle $ & 
    $\Gamma^d_N = \langle \Gamma^{d \, {\rm (A+B+C)}} \rangle $ 
    & 
    $\Gamma^d_N = \Gamma^{d,p \,{\rm (A+B+C)}} (k)$ 
    \\
    \hline
    $Y_B$ ({\small 10 GeV}) & 
    $3.8 \times 10^{-10}$ & 
    $2.7 \times 10^{-10}$ &
    $2.9 \times 10^{-10}$
    \\
    $Y_B$ ({\small 100 MeV}) & 
    $5.8 \times 10^{-14}$ & 
    $3.9 \times 10^{-14}$ &
    $5.4 \times 10^{-14}$
    \\
    \hline
  \end{tabular}
  \caption{\it 
    Comparison of $Y_B$ for cases I, II and III
    when $M_N =$ 10 GeV and 100 MeV. 
    See the details in the text.
  }
  \label{tab:YB}
\end{table}
%%%%%%%%%%%%%%%%%%%%%%%%%%%%%%%%%%%%%%%%%%%%%%%%%%%%%%%%%%%%%%%%%%%%
%
%%%%%%%%%%%%%%%%%%%%%%%%%%%%%%%%%%%%%%%%%%%%%%%%%%%%%%%%%%%%%%%%%%%%
%%%%% ** Figure ** %%%%%%%%%%%%%%%%%%%%%%%%%%%%%%%%%%%%%%%%%%%%%%%%%
\begin{figure}[t]
  \centerline{
  \includegraphics[scale=1.5]{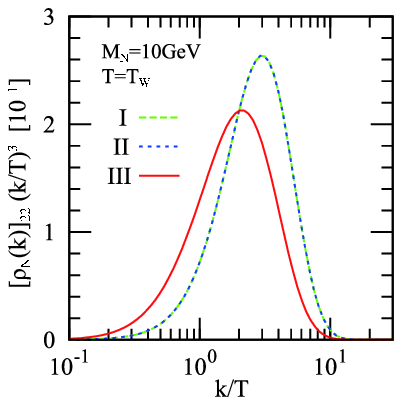}%
  \hspace{1cm}
  \includegraphics[scale=1.5]{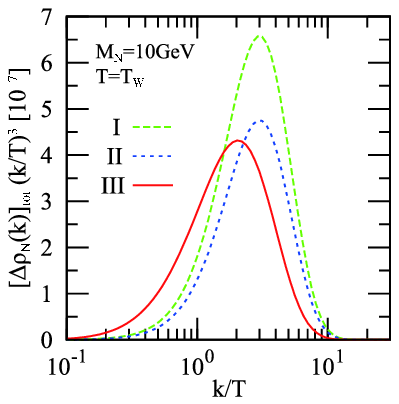}%
  }%
  \centerline{
  \includegraphics[scale=1.5]{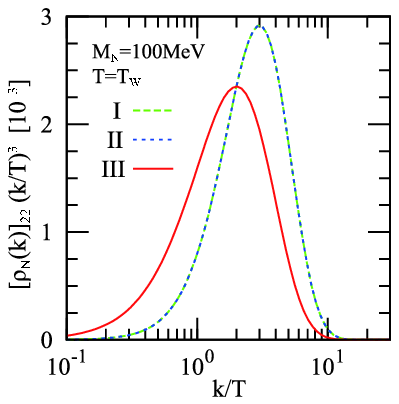}%
  \hspace{1cm}
  \includegraphics[scale=1.5]{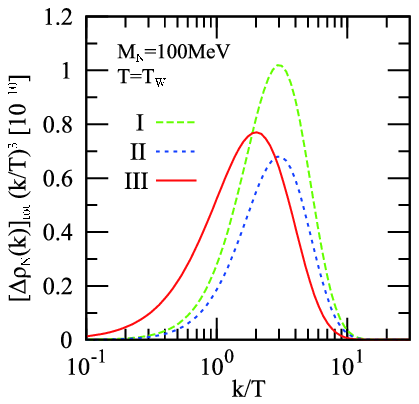}%
  }%
  \caption{\it Comparison of momentum distributions of 
    $[\rho_N (k)]_{22}$ (left panel) and
    $[\Delta \rho_N(k)]_{\rm tot}$ (right panel)
    for the cases I, II and III in Tab.~\ref{tab:YB}.  
    The distributions are evaluated at $T=T_W$
    and are multiplied by $(k/T)^3$.  Here we take $M_N=$10 GeV and 
    100 MeV.  }
  \label{fig:COMP}
\end{figure}
%%%%%%%%%%%%%%%%%%%%%%%%%%%%%%%%%%%%%%%%%%%%%%%%%%%%%%%%%%%%%%%%%%%%

Having specified the differences between
various sets of the kinetic equations to now,
we can compare quantitatively the yields of the BAU.
The results are summarized in Tab.~\ref{tab:YB},
where we take the same parameter set described in
Sec.~\ref{sec:NS}, but both $M_N = $ 10 GeV and 100 MeV are
considered.  The cases I and II correspond to Ref.~\cite{Asaka:2010kk}
and to Eqs.~(\ref{eq:KEAP_MU}) and (\ref{eq:KEAP_RN}), respectively.
For comparison, we also show the results from Sec.~\ref{sec:NS} as the
case III.

Let us first make the comparison between the cases I and II in which
the momentum dependence in $\rho_N$ is approximately included.  The
momentum distributions of $[\rho_{N}(k)]_{22}$ and $[\Delta
\rho_{N}(k)]_{\rm tot}$ are shown in Fig.~\ref{fig:COMP}.  Since the
momentum dependence of $\rho_N$ is taken as (\ref{eq:AP_RHON}), only
the normalization is important.  As seen in the occupation numbers,
the yields of $N_2$ (and $N_3$) are almost the same because the
production dominantly occurs through the second term in
Eq.~(\ref{eq:KEAP_RN}) which is common for both cases.  On the other
hand, the asymmetries are mainly generated through the communication
terms of the kinetic equations, and then $Y_B$ can vary depending on
the structure of these terms.  We find that the BAU in the case II is
smaller than I by $\sim 2/3$, and accordingly the distributions of
$[\Delta \rho_{N}(k)]_{\rm tot}$ are different by the same amount
since $Y_B \propto Y_{\Delta N_{\rm tot}} (= - Y_{\Delta L_{\rm
    tot}})$.

The origin of the factor $2/3$ can be understood as follows: Active
flavor asymmetries $Y_{\Delta L_\alpha}$ are almost the same between
two cases.  This is because the three scattering processes equally
contribute to the generation (as long as chemical potentials are small
enough).  On the other hand, only the processes (A) and (B) give a
dominant contribution to $Y_{\Delta L_{\rm tot}}$ and $Y_{\Delta
  N_{\rm tot}}$ through the third term in Eq.~(\ref{eq:KEAP_RN}).  The
contribution from the process (C) is suppressed by ${\cal O}(F^2)$.
The difference in $Y_B$ is just coming from the change of the
coefficient of this term.  This is the impact of the new communication
terms found in this analysis.

Finally, we turn to discuss the difference between the cases II and
III in order to estimate the importance of the momentum dependence in
$\rho_N$.  Note that the difference lies only in the treatment of the
momentum dependence between the two cases.  When we take the smaller
value of Majorana mass as $M_N = 100$ MeV neutrino Yukawa coupling
constants become sufficiently small that all the modes of interest can
be away from the equilibrium one.  As mentioned in Sec.~\ref{sec:NS},
the momentum dependence in the rates boost (or disturb) the
production of the occupation numbers and also the asymmetries for the
modes with lower (or higher) momenta.  These behaviors can be seen in
the distributions in Fig.~\ref{fig:COMP}.  In this case $Y_B$ is
enhanced by about 40\% due to the inclusion of the momentum
dependence.

On the other hand, when $M_N = $ 10 GeV, the large Yukawa coupling
constants lead to another effect.  The low momentum modes are likely
to gets in thermal equilibrium and the matrix of densities becomes
$\rho_N \to \rho^{\rm eq} \mathbf{1}$.  The asymmetries carried by
such modes receive the wash-out effect.
This point is clearly seen by comparing the distributions
$[\Delta \rho_N (k)]_{\rm tot}$ of the cases II and III.
As a result, the enhancement factor of $Y_B$ is reduced to
about 10\%.

%%%%%%%%%%%%%%%%%%%%%%%%%%%%%%%%%%%%%%%%%%%%%%%%%%%%%%%%%%%%%%%%%%%%
%%%%%%%%%%%%%%%%%%%%%%%%%%%%%%%%%%%%%%%%%%%%%%%%%%%%%%%%%%%%%%%%%%%%
\section{Conclusions}
\label{sec:Conc}
%%%%%%%%%%%%%%%%%%%%%%%%%%%%%%%%%%%%%%%%%%%%%%%%%%%%%%%%%%%%%%%%%%%%
%%%%%%%%%%%%%%%%%%%%%%%%%%%%%%%%%%%%%%%%%%%%%%%%%%%%%%%%%%%%%%%%%%%%
In this paper we have investigated baryogenesis induced by the flavor
oscillation between sterile (right-handed) neutrinos $N_2$ and $N_3$
in the $\nu$MSM.  We have presented the kinetic equations for the
matrices of densities of sterile neutrinos $\rho_N$ and $\rho_{\bar
  N}$ and chemical potentials of active (left-handed) leptons
$\mu_{\nu_\alpha}$ shown in Eqs.~(\ref{eq:KE1}) and (\ref{eq:KE3}).
By using these equations the time evolution for the matrices of
densities can be traced for each mode, and the momentum distributions
can be found at any moments relevant for baryogenesis.  As a result
they allow us to estimate the yield of the baryon asymmetry more
precisely.

We have evaluated the destruction and production rates of sterile and
active leptons for the kinetic equations.  Especially, we have paid a
special attention to the following two respects.  The first is to
include the momentum dependence of the rate correctly, which is
mandatory to obtain the equations of $\rho_{N, \bar N}$ for a given
momentum.  The second is to include the effect from the deviation of
the initial or final state from the thermal equilibrium in the
scattering process.  As explained in Sec.~\ref{sec:KE}, 
the second point is crucially important in the construction of 
the kinetic equations for baryogenesis in the $\nu$MSM.  
We have shown that the terms connecting between sterile and active sectors,
which are required for the successful baryogenesis, arise in a consistent
way from the second effect to the rate.

To be concrete, the terms connecting the matrices of densities
$\rho_N$ and $\rho_L$, which introduced in Ref.~\cite{Asaka:2005pn},
appear as the corrections to the production rates $\delta \Gamma_N^p$
and $\delta \Gamma_\nu^p$ via the scattering processes (A) and (B).
As shown in Sec.~\ref{sec:Comparison} the prefactor is different by
2/3 to the previous result in Ref.~\cite{Asaka:2010kk}.  Moreover, we
have found the new terms which link $\rho_N$ to $\rho_{\bar L}$
(rather than $\rho_L$).  They are induced as the corrections $\delta
\Gamma_N^d$ and $\delta \Gamma_\nu^d$ to the destruction rates via the
process (C).  Importantly, such a term is proportional to the product
of $\rho_{N , {\bar N}}$ and $\mu_{\nu_{\alpha}}$.  Thus, our kinetic
equations are no longer linear in $\rho_N , \rho_{\bar N}$ and
$\mu_{\nu_{\alpha}}$.  It should also be stressed that both types of
these terms are required to ensure the conservation of the total
lepton number.  The impacts of these terms on the generation of the
baryon asymmetry have been described in detail.

We have then investigated the numerical solutions of the kinetic
equations.  Because of the momentum dependence in the rates the
production of sterile neutrinos is enhanced for the modes with $k/T
\lesssim 1$ while diminished for the higher momentum modes.
Consequently, the momentum distributions of $\rho_N$ and $\rho_{\bar
  N}$ are significantly distorted from the equilibrium one.  This
clearly shows the importance to utilize the momentum-dependent kinetic
equations.  It has also been found that the low modes get in the
thermal equilibrium and the asymmetries carried by them receive the
wash-out effect at higher temperatures.  In our choice of the
parameters, the mode with $k/T \simeq 2$ is the most significant for
the occupation numbers as well as the asymmetries including the BAU.

The comparison with the previous works has also been discussed.  We
have first explained what kind of approximations are needed to
reproduce the previous kinetic equations staring from the present
ones.  It is shown that the use of the destruction
rate~\cite{Akhmedov:1998qx} obtained by the thermal average is crucial
to ensure the lepton number conservation when we treat the momentum
dependence of $\rho_N$ approximately.  We have then compare
quantitatively the yields of the BAU.  Apart from the
correct treatment of the momentum dependence, the present equations
gives the smaller BAU by a factor of $\simeq 2/3$ compared with
Ref.~\cite{Asaka:2010kk}.  Within the specific
choice of parameters, the inclusion of the momentum dependence
enhances the BAU by about 40\% and 10\% for $M_N=100$ MeV and 10 GeV,
respectively.  This is because the enhancement in the generation of
the asymmetry from the lower modes overcomes the suppression induced by
the higher modes. On the other hand, such an amplification
may be disturbed by the wash-out effect for large neutrino Yukawa
coupling constants.

The present analysis shows that the momentum distribution of the
matrices of densities differs from the equilibrium one
and the inclusion of  their momentum dependence modifies
the yield of the BAU in non-trivial manner depending on the
parameters of neutrino Yukawa coupling constants and masses
of sterile neutrinos.  Therefore, it is important to explore
the full parameter space of the $\nu$MSM accounting for 
the observed BAU by using the kinetic equations presented
in this paper.  This issue will be performed elsewhere.

%%%%%%%%%%%%%%%%%%%%%%%%%%%%%%%%%%%%%%%%%%%%%%%%%%%%%%%%%%%%%%%%%%%%
%%%%%%%%%%%%%%%%%%%%%%%%%%%%%%%%%%%%%%%%%%%%%%%%%%%%%%%%%%%%%%%%%%%%
\section*{Acknowledgments}
%%%%%%%%%%%%%%%%%%%%%%%%%%%%%%%%%%%%%%%%%%%%%%%%%%%%%%%%%%%%%%%%%%%%
%%%%%%%%%%%%%%%%%%%%%%%%%%%%%%%%%%%%%%%%%%%%%%%%%%%%%%%%%%%%%%%%%%%%
We would like to thank E. Kh. Akhmedov, A. Kartavtsev and M. Shaposhnikov
for valuable discussions and comments,
and also to Particle and
Astroparticle Division of Max-Planck-Institut f\"ur Kernphysik at
Heidelberg for hospitality.  
The work of T.A. was supported by KAKENHI (No.~21540260) in JSPS.
T.A. and S.E. are supported from Strategic Young Researcher Overseas Visits
Program for Accelerating Brain Circulation in JSPS.

%%%%%%%%%%%%%%%%%%%%%%%%%%%%%%%%%%%%%%%%%%%%%%%%%%%%%%%%%%%%%%%%%%%%
%%%%%%%%%%%%%%%%%%%%%%%%%%%%%%%%%%%%%%%%%%%%%%%%%%%%%%%%%%%%%%%%%%%%
%%%%% ** Reference ** %%%%%%%%%%%%%%%%%%%%%%%%%%%%%%%%%%%%%%%%%%%%%%
%%%%%%%%%%%%%%%%%%%%%%%%%%%%%%%%%%%%%%%%%%%%%%%%%%%%%%%%%%%%%%%%%%%%
%%%%%%%%%%%%%%%%%%%%%%%%%%%%%%%%%%%%%%%%%%%%%%%%%%%%%%%%%%%%%%%%%%%%

%%%%%%%%%%%%%%%%%%%%%%%%%%%%%%%%%%%%%%%%%%%%%%%%%%%%%%%%%%%%%%%%%%%%
%%%%%%%%%%%%%%%%%%%%%%%%%%%%%%%%%%%%%%%%%%%%%%%%%%%%%%%%%%%%%%%%%%%%
%%%%%%%%%%%%%%%%%%%%%%%%%%%%%%%%%%%%%%%%%%%%%%%%%%%%%%%%%%%%%%%%%%%%
%%%%%%%%%%%%%%%%%%%%%%%%%%%%%%%%%%%%%%%%%%%%%%%%%%%%%%%%%%%%%%%%%%%%

\begin{thebibliography}{100}

%%%%%%%%%%%%%%%%
\bibitem{Riotto:1999yt}
  A.~Riotto and M.~Trodden,
  %``Recent progress in baryogenesis,''
  Ann.\ Rev.\ Nucl.\ Part.\ Sci.\  {\bf 49} (1999) 35
  [arXiv:hep-ph/9901362].
  %%CITATION = ARNUA,49,35;%%\bibitem{Seesaw}

\bibitem{Fukugita:1986hr}
  M.~Fukugita and T.~Yanagida,
  %``Baryogenesis Without Grand Unification,''
  Phys.\ Lett.\  B {\bf 174} (1986) 45 .
  %%CITATION = PHLTA,B174,45;%%

\bibitem{Seesaw}
%\cite{Minkowski:1977sc}
%\bibitem{Minkowski:1977sc}
P.~Minkowski,
%``Mu $\to$ E Gamma At A Rate Of One Out Of 1-Billion Muon Decays?,''
Phys.\ Lett.\ B {\bf 67} (1977) 421;
%%CITATION = PHLTA,B67,421;%%
%%%%
T.~Yanagida,
in {\em Proc. of the Workshop on the Unified Theory
and the Baryon Number in the Universe}, 
Tsukuba, Japan, Feb.~13-14, 1979, p.~95, 
eds. O.~Sawada and S.~Sugamoto, 
(KEK Report KEK-79-18, 1979, Tsukuba); 
Progr.\ Theor.\ Phys.\ {\bf 64} (1980) 1103 ; 
%%%%
M.~Gell-Mann, P.~Ramond and R.~Slansky, 
in {\em Supergravity}, 
eds. P.~van~Niewenhuizen and D.~Z.~Freedman
(North Holland, Amsterdam 1980);
%%%%
P.~Ramond, 
in {\em Talk given at the Sanibel Symposium}, 
Palm Coast, Fla., Feb.~25-Mar.~2, 1979, preprint CALT-68-709
(retroprinted as hep-ph/9809459);
%%%%
S.~L.~Glashow,
in {\em Proc. of the Carg\'ese  Summer Institute on Quarks and Leptons},
Carg\'ese, July 9-29, 1979, 
eds. M.~L\'evy et. al, , (Plenum, 1980, New York), p707.


\bibitem{Akhmedov:1998qx}
  E.~K.~Akhmedov, V.~A.~Rubakov and A.~Y.~Smirnov,
  %``Baryogenesis via neutrino oscillations,''
  Phys.\ Rev.\ Lett.\  {\bf 81} (1998) 1359
  [arXiv:hep-ph/9803255].
  %%CITATION = PRLTA,81,1359;%%  

\bibitem{Kuzmin:1985mm}
  V.~A.~Kuzmin, V.~A.~Rubakov and M.~E.~Shaposhnikov,
  %``On The Anomalous Electroweak Baryon Number Nonconservation In The Early
  %Universe,''
  Phys.\ Lett.\  B {\bf 155} (1985) 36.
  %%CITATION = PHLTA,B155,36;%%

\bibitem{Asaka:2005an}
  T.~Asaka, S.~Blanchet and M.~Shaposhnikov,
  %``The nuMSM, dark matter and neutrino masses,''
  Phys.\ Lett.\  B {\bf 631} (2005) 151.
%  [arXiv:hep-ph/0503065].
  %%CITATION = PHLTA,B631,151;%%

\bibitem{Asaka:2005pn}
  T.~Asaka and M.~Shaposhnikov,
  %``The nuMSM, dark matter and baryon asymmetry of the universe,''
  Phys.\ Lett.\  B {\bf 620} (2005) 17.
%  [arXiv:hep-ph/0505013].
  %%CITATION = PHLTA,B620,17;%%

\bibitem{Boyarsky:2009ix}
  A.~Boyarsky, O.~Ruchayskiy and M.~Shaposhnikov,
  %``The role of sterile neutrinos in cosmology and astrophysics,''
  Ann.\ Rev.\ Nucl.\ Part.\ Sci.\  {\bf 59} (2009) 191
  [arXiv:0901.0011 [hep-ph]].
  %%CITATION = ARNUA,59,191;%%

\bibitem{Bezrukov:2007ep}
  F.~L.~Bezrukov and M.~Shaposhnikov,
  %``The Standard Model Higgs boson as the inflaton,''
  Phys.\ Lett.\  B {\bf 659} (2008) 703
  [arXiv:0710.3755 [hep-th]].
  %%CITATION = PHLTA,B659,703;%%
  

\bibitem{Shaposhnikov:2008pf}
  M.~Shaposhnikov,
  %``The nuMSM, leptonic asymmetries, and properties of singlet fermions,''
  JHEP {\bf 0808} (2008) 008
  [arXiv:0804.4542 [hep-ph]].
  %%CITATION = JHEPA,0808,008;%%
  

\bibitem{Asaka:2010kk}
  T.~Asaka and H.~Ishida,
  %``Flavour Mixing of Neutrinos and Baryon Asymmetry of the Universe,''
  Phys.\ Lett.\  B {\bf 692} (2010) 105
  [arXiv:1004.5491 [hep-ph]].
  %%CITATION = PHLTA,B692,105;%%
 

\bibitem{Canetti:2010aw}
  L.~Canetti and M.~Shaposhnikov,
  %``Baryon Asymmetry of the Universe in the NuMSM,''
  JCAP {\bf 1009} (2010) 001
  [arXiv:1006.0133 [hep-ph]].
  %%CITATION = JCAPA,1009,001;%%

\bibitem{Dolgov:1980cq}
  A.~D.~Dolgov,
  %``Neutrinos In The Early Universe,''
  Sov.\ J.\ Nucl.\ Phys.\  {\bf 33} (1981) 700
  [Yad.\ Fiz.\  {\bf 33} (1981) 1309].
  %%CITATION = YAFIA,33,1309;%%
\bibitem{Barbieri:1990vx}
  R.~Barbieri and A.~Dolgov,
  %``Neutrino oscillations in the early universe,''
  Nucl.\ Phys.\  B {\bf 349} (1991) 743.
  %%CITATION = NUPHA,B349,743;%%
\bibitem{Sigl:1992fn}
  G.~Sigl and G.~Raffelt,
  %``General kinetic description of relativistic mixed neutrinos,''
  Nucl.\ Phys.\  B {\bf 406} (1993) 423.
  %%CITATION = NUPHA,B406,423;%%

%\cite{Sakharov:1967dj}
\bibitem{Sakharov:1967dj}
  A.~D.~Sakharov,
  %``Violation of CP Invariance, c Asymmetry, and Baryon Asymmetry of the
  %Universe,''
  Pisma Zh.\ Eksp.\ Teor.\ Fiz.\  {\bf 5} (1967) 32
  [JETP Lett.\  {\bf 5} (1967\ SOPUA,34,392-393.1991\ UFNAA,161,61-64.1991) 24].
  %%CITATION = UFNAA,161,NO.561;%%

\bibitem{Dasgupta:2009mg}
  B.~Dasgupta, A.~Dighe, G.~G.~Raffelt and A.~Y.~Smirnov,
  %``Multiple Spectral Splits of Supernova Neutrinos,''
  Phys.\ Rev.\ Lett.\  {\bf 103} (2009) 051105
  [arXiv:0904.3542 [hep-ph]].
  %%CITATION = PRLTA,103,051105;%%

\bibitem{Duan:2010bg}
  H.~Duan, G.~M.~Fuller, Y.~-Z.~Qian,
  %``Collective Neutrino Oscillations,''
  Ann.\ Rev.\ Nucl.\ Part.\ Sci.\  {\bf 60 } (2010)  569-594.
  [arXiv:1001.2799 [hep-ph]].

\bibitem{Weldon:1982bn}
  H.~A.~Weldon,
  %``Effective Fermion Masses Of Order Gt In High Temperature Gauge Theories
  %With Exact Chiral Invariance,''
  Phys.\ Rev.\  D {\bf 26} (1982) 2789.
  %%CITATION = PHRVA,D26,2789;%%


\bibitem{arXiv:1106.6028}
  G.~L.~Fogli, E.~Lisi, A.~Marrone, A.~Palazzo and A.~M.~Rotunno,
  %``Evidence of $\theta_{13}$>0 from global neutrino data analysis,''
  Phys.\ Rev.\ D\ {\bf 84} (2011) 053007
  [arXiv:1106.6028 [hep-ph]].
  %%CITATION = PHRVA,D84,053007;%%

\bibitem{hep-ph/0511246}
  Y.~Burnier, M.~Laine and M.~Shaposhnikov,
  %``Baryon and lepton number violation rates across the electroweak crossover,''  
  JCAP\ {\bf 0602} (2006) 007  [hep-ph/0511246].  
%%CITATION = JCAPA,0602,007;%%

\bibitem{Asaka:2011pb}
  T.~Asaka, S.~Eijima and H.~Ishida,
  %``Mixing of Active and Sterile Neutrinos,''
  JHEP {\bf 1104} (2011) 011
  [arXiv:1101.1382 [hep-ph]].
  %%CITATION = JHEPA,1104,011;%%

\end{thebibliography}
\end{document}